\newif\ifconfver
\confvertrue        %declaring conference version true

\newif\ifcutshort      %this level shortens the equations
\cutshorttrue

\newif\ifcutshortlvltwo  %this level takes out some examples, figs., and sim.
\cutshortlvltwofalse

\ifconfver
\documentclass[10pt,twocolumn,twoside]{IEEEtran}
\else
\documentclass[11pt,draftcls,onecolumn,draftclsnofoot]{IEEEtran}
\fi

\usepackage{cite}
\usepackage{graphicx}
\usepackage{psfrag}
\usepackage{url}
\usepackage{stfloats}
\usepackage{amsfonts,amssymb,amsmath,bm,paralist,theorem,cite,ifthen,color}
\usepackage{array}
\usepackage{caption}
\usepackage{calc}
\usepackage{subfigure}
\usepackage{longtable}
\usepackage{stfloats}  % Written by Sigitas Tolusis
\usepackage{graphics,booktabs,color,epsfig,subfigure,enumerate}
\usepackage{multirow}
\usepackage{algorithm,algorithmic}
\usepackage{epstopdf}
\usepackage{array}
\usepackage{stfloats}
\usepackage{mathrsfs}
\newtheorem{Lemma}{Lemma}
\newtheorem{Claim}{Claim}

\newtheorem{Theorem}{Theorem}

\theorembodyfont{\rmfamily}

{\begin{list}{}{
			\settowidth{\labelwidth}{\mbox{\textnormal{#1}}}%
			\setlength{\leftmargin}{\labelwidth+\labelsep}}}%
	{\end{list}}

\begin{document}
	
	\bibliographystyle{IEEEtran}

	\def\blue{\color{blue}}
	\def\red{\color{red}}
	\definecolor{orange}{RGB}{255,107,0}
	\def\orange{\color{orange}}
	
	% ==================================================
	\title{Joint Secure Communication and Radar Beamforming: A Secrecy-Estimation Rate-Based Design}
	
	\ifconfver \else {\linespread{1.1} \rm \fi
		\author{ Rong Wen, Ying Zhang, Qiang Li and  Youxi Tang
			\thanks{R. Wen and Y. Tang are with the National Key Laboratory of Science and Technology on Communications, University of Electronic Science and Technology of China, Chengdu 611731, China.

				Q. Li ({\it Corresponding author}) and Y. Zhang are with
				School of Information and Communication  Engineering, University of Electronic Science and Technology of China, Chengdu, P.~R.~China, 611731. E-mail: lq@uestc.edu.cn} }

			%\thanks{X. Wu is with College of Electronics and Information Engineering, Shenzhen University, Shenzhen 518060, China. E-mail:xxwu.eesissi@szu.edu.cn}
			%	\thanks{Part of this work has been submitted to the IEEE International Conference
			%		on Acoustics, Speech, and Signal Processing (ICASSP) 2022 for possible
			%		publication~\cite{qli22}.}
%		}
		%\author{ Qiang Li$^\dagger$, Ying Zhang, Xiaoxiao Wu and Jing Ran Lin
		%	\thanks{Q. Li, Y. Zhang and J. Lin are with
		%		School of Information and Communication  Engineering, University of Electronic Science and Technology of China, Chengdu, P.~R.~China, 611731. E-mail: lq@uestc.edu.cn}
		%		\thanks{X. Wu is with College of Electronics and Information Engineering, Shenzhen University, Shenzhen 518060, China. E-mail:xxwu.eesissi@szu.edu.cn}
		%\thanks{$^\dagger$Corresponding author.}
		%}
		\maketitle

% Abstract
\begin{abstract}
This paper considers transmit beamforming in dual-function radar-communication (DFRC) system, where a DFRC transmitter simultaneously communicates with a communication user and detects a malicious target with the same waveform. Since the waveform is embedded with information, the information is risked to be intercepted by the target. To address this problem, physical-layer security technique is exploited. By using secrecy rate and estimation rate as performance measure for communication and radar, respectively, three secrecy rate maximization (SRM) problems are formulated, including the SRM with and without artificial noise (AN), and robust SRM. For the SRM beamforming, we prove that the optimal beamformer can be computed in closed form. For the AN-aided SRM, by leveraging alternating optimization similar closed-form solution is obtained for the beamformer and the AN covariance matrix. Finally, the imperfect CSI of the target is also considered under the premise of a moment-based random phase-error model on the direction of arrival at the target.
% By using the recent advances in distributionally robust optimization and the semidefinite relaxation, we derive a tractable solution to the robust SRM. Simulation results demonstrate the efficacy and robustness of the proposed designs.
Simulation results demonstrate the efficacy and robustness of the proposed designs.

\end{abstract}


\ifconfver \else \IEEEpeerreviewmaketitle} \fi
%
%
%
%% ==================================================
\section{Introduction}
Traditionally, communication and radar  are developed independently due to different objectives and application scenarios. However, recently there is growing interest in integrating communication and radar functions within one platform~\cite{Liufan20,Niyato}, which results in joint communication and radar (JCR). The research of JCR is  motivated by at least the following two folds. Firstly, with the ever increasing demand of high-speed communication has driven the communication to higher frequency, e.g., mmWave, which overlaps with the conventional radar frequency; the convergence trend of the communication and radar frequency makes it possible to process both signals within one platform. Secondly, there are
emerging applications, e.g.,  autonomous vehicle system, flying wireless mesh
networks, involving both sensing and communication. By merging  the two functions as a whole, one can achieve  a more efficient system design --- communication can better adapt to the environment with sensing, and meanwhile, sensing accuracy can be improved with information exchange from communication. 

The development of JCR roughly consists of three stages, namely, partial hardware sharing, coexistence of radar and communication (CRC), and dual-function radar-communication (DFRC) system. Back to 1980s, the U.S. Air Force had launched the ``Pave Pillar'' program, whose goal is to integrate radar, communication and  electronic warfare functions by sharing part of the hardware, so that the size, weight and power consumption of the system can be reduced as compared with the separate architecture. However, this integration is largely at the hardware-reuse level and the co-channel interference between radar and communication still exists. Later, with the  advances of signal processing, the concept of CRC is developed in order to alleviate the interference between radar and communication. The  idea of CRC is similar to cognitive radio --- the communication and the radar cooperatively share the spectrum with controllable interference management strategies, e.g., spatial nulling~\cite{Goldsmith13}. The CRC allows to use the spectrum in a more efficient way, but the communication and radar functions are still isolated, at least at the waveform level. More recently, DFRC is seen as the most advanced stage of JCR, which deeply integrates the communication and radar functions with the same hardware and software, and both functions can be simultaneously realized with a single waveform~\cite{Niyato}. Due to the unified architecture and high spectral efficiency, DFRC has gained considerable attention~\cite{Niyato,Goldsmith13,Liufan20,Chambers18,Chenguang20,Nanchi21,Mashuai19,Athina17,Bliss14,Liufan18,Batu18} and is also seen as one of the promising technologies in 6G communications to realize integration of communication and sensing (ISAC) applications~\cite{Tan21,Wei22}.

In this work, we will focus on the transmit signal optimization for DFRC, with an emphasis on the information security. In the existing literature, there have been a pile of works investigating DFRC from different perspective, including the waveform design, resource allocation, joint beamforming and capacity characterization, to name a few; see the recent survey papers~\cite{Liufan20,Niyato} for the details. However, the information security aspect of DFRC is relatively less investigated, except for a few recent works~\cite{ShuaiMa_twc,Chambers18,Nanchi21,Mashuai19,Batu18,Su21_arxiv}. Since DFRC employs a single waveform to simultaneously detect the target and send information to the communication user (CU), the information embedded in the waveform has a risk to leak to the malicious target. This information leakage could be more serious in the DFRC applications. Specifically, consider the target is also an eavesdropper and intends to intercept the information. For the radar purpose, it requires the DFRC transmitter to focus the energy of the transmit signal on the target, say by transmit beamforming, in order to improve the estimation accuracy. However, this in turn exposes the  waveform to and favors interception at the malicious target.  
To circumvent the above difficulty, physical-layer security (PLS) technique~\cite{Liang_book} has recently been employed to secure DFRC~\cite{Wei22}. PLS is an information-theoretic approach to achieve confidentiality at the physical layer. The origin of PLS is due to Wyner's seminal work~\cite{Wyner1975} in 1970s, where the author showed that perfect secrecy can be achieved at the physical layer for degraded wiretap channel. Later, the idea of PLS was generalized to non-degraded channel by Csisz$\acute{\text{a}}$r and K$\ddot{{\text o}}$rner~\cite{Csiszar1978}, and more recently the multi-input multi-output (MIMO) Gaussian channel~\cite{Oggier2008}. Unlike the upper layer encryption approach, PLS guarantees that the legitimate receiver can correctly decode the information, and meanwhile the eavesdropper cannot retrieve any useful information from his observation. Back to the DFRC application, with the aid of PLS, even if the malicious target intercepts the DFRC waveform, he still cannot obtain any information encoded in the waveform.

Following the idea of PLS, we consider a transmit beamforming optimization for DFRC, where a multi-antenna DFRC transmitter aims to send information to a CU, and meanwhile measure the distance parameter of a malicious target from echo. For the communication aspect, we model the DFRC transmitter, the CU and the target as three nodes wiretap channel, and leverage on the \emph{secrecy rate}---the maximum information rate at which perfect secrecy can be attained~\cite{Liang_book}---to measure the communication performance. While for the radar aspect, we adopt the \textit{estimation rate}---the minimum number of bits that need to be used to encode the Kalman residual~\cite{Bliss17}---to measure the radar performance. We should mention that estimation rate was originally introduced by Bliss in~\cite{Bliss14} to unify the performance measure of communication and radar from the view of information theory. Upon the above model and performance measures, we aim at maximizing the secrecy rate at the CU, and meanwhile satisfying a pre-specified estimation rate for the target and the average transmit power budget at the DFRC transmitter. This secrecy rate maximization (SRM) problem is nonconvex by nature, due to the nonconvex secrecy rate function. Nevertheless, we first show that the nonconvex SRM problem is actually hidden convex and can be optimally solved via semidefinite relaxation (SDR) and rank reduction~\cite{Huang2009}. Then, by inspecting the optimal structure of the beamformer, we show that the SRM problem can be reformulated as a single-variable optimization problem with box constraint, whose optimal solution can be obtained in closed form via computing the roots of a quadratic equation.

In the traditional PLS context, it is well-known that artificial noise (AN)
is effective in improving security by proactively sending noise to jam the eavesdropper~\cite{Goel,qli13}. As for DFRC, it is envisioned that AN is not only beneficial for securing information, it is also helpful for radar because for the
	DFRC transmitter the AN is deterministically known as a prior, and its echo can be	further exploited by the DFRC transmitter for the radar purpose~\cite{Chambers18,Nanchi21,Mashuai19}. In light of this, we also consider an AN-aided DFRC beamforming for the SRM problem. The resultant problem is more challenging, owing to the coupled AN and beamformer. We employ an alternating optimization (AO) approach to alternately optimize the beamformer and the AN. For the beamformer, the previously established closed-form solution can be directly applied. For the AN, we again show that it can be solved with a close-form solution via computing the roots of a quadratic equation.

It should be noted that in the above SRM problems we have implicitly assumed that the channel, more specifically the direction under the planar array model, of the target is accurately known. However, in practice there could be some uncertainty on the direction due to inaccurate estimation. To account for this, we further consider a robust AN-aided SRM problem by assuming that  the direction of the target is randomly distributed. The exact distribution of the direction is not known, except for its first- and second-order moments. The reason for considering the moment-based random model is due to the fact that in practice it is relatively easy to estimate the mean and covariance of the direction  rather than the complete distribution from initial radar searching stage. Under this moment-based uncertainty model, we formulate a  distributionally robust SRM (DR-SRM) problem to maximize the outage secrecy rate subject to the outage probability constraint, which is evaluated with respect to (w.r.t.) any distribution fulfilling the given first- and second-order moments,  on the target's estimation rate and the total power budget. We should mention that this DR-SRM is different from the conventional worst-case robust model~\cite{Nanchi21} or the Gaussian random error model~\cite{So16}. Without assuming specific distribution on the direction, the resultant robust design is able to provide maximum robustness to the mismatch of the distribution, which is particular important for secrecy applications due to non-cooperation of the eavesdropper. The DR-SRM is in essence a semi-infinite chance-constraint problem. By exploiting recent advances on distributionally robust optimization~\cite{Zymler,Ye}, we show that after some approximation, the DR-SRM problem can be recast as a conic optimization problem, which can be tackled by the SDR technique and the Charnes-Cooper transformation~\cite{Charnes1962}.

There are some related works worth mentioning. In~\cite{Chambers18}, the authors first considered the PLS in DFRC with a DFRC transmitter, a communication user and a target. Both SRM at the communication user and signal-to-interference-plus-noise ratio (SINR) maximization at the target are considered. By exploiting Taylor series approximation, iterative methods are proposed to solve the both problems. In~\cite{Nanchi21}, Su~{\it et~al.} considered AN-aided eavesdropper's SNR minimization with desired beampattern and SINR at the legitimate receiver. Both perfect CSI and imperfect CSI are considered. By exploiting SDR and $\cal S$-procedure, the authors proposed various iterative methods to optimize the beamformer and the AN. We should mention that the robust model in~\cite{Nanchi21} is the worst-case error model, which is different from the chance-constrained model in our work. In~\cite{Batu18}, a bistatic DFRC is considered. Different from~\cite{Chambers18,Nanchi21}, the authors considered the DFRC transmitter simultaneously sends the radar waveform and the information signal in overlapping or non-overlapping manner. The radar receiver, instead of the target is treated as potential eavesdropper. Under such setting, they aim at  maximizing the SINR at the radar receiver while providing pre-specified  secrecy rate for the communication user. AO with iterative SDP method is proposed to handle the SINR maximization problem. In~\cite{Mashuai19}, the authors considered joint AN and precoding matrix design for DFRC when there are multiple targets. A max-min SRM problem is considered with SNR threshold at the target. By exploiting a conjugate reformulation of the log-det function, a three block-coordinate ascent (BCA) algorithm is proposed. In~\cite{Su21_arxiv}, the authors considered using a directional modulation approach to secure DFRC and proposed a fractional programming algorithm to solve the radar SINR maximization problem. In~\cite{ShuaiMa_twc}, the authors considered a covert communication under the radar probing waveforms; a joint design of the target detection beamformer and communication beamformer was proposed. It should also be noted that besides PLS, there are also some other works investigating privacy in DFRC, e.g.,~\cite{Athina17,Athina17_2}.  To summarize, compared with the existing works, the  contribution of this work is summarized as follows.
\begin{enumerate}
\item  The existing works on secure DFRC mainly adopt the  secrecy rate and the SINR as performance measure, whereas  this work  studies the secure DFRC problem from an information-theoretic perspective by proposing a new  secrecy-estimation rate-based design formulation. 
\item Under the premise of perfect channel state information (CSI),  the secure DFRC beamformer designs with and without AN are developed. By analyzing the problem structure, we identify the optimal beamformer and AN covariance structure and establish both closed-form and semi-closed-form solutions for the secure DFRC  design problems. Thanks to the closed-form solution, the proposed designs can be efficiently computed without calling a general optimization solver as most existing works did.
\item This work further develops a robust solution for the secure DFRC  design problem under imperfect CSI case. A  distributionally robust chance-constrained model is proposed to make the resultant design robust to any random  channel error with pre-specified the first- and second-order moments. The distributionally robust DFRC model is new and different from the  existing robust model, e.g.~\cite{Nanchi21}. 
\end{enumerate}

%\subsection{Organization and Notations}
The remainder of this paper is organized as follows. Section~\ref{sec:model} introduces the system model and problem formulation. Sections~\ref{sec:solution_no_AN} and \ref{sec:solution_with_AN} develop closed-form solutions to the SRM problem and the AN-aided SRM problem, respectively. Section~\ref{sec:robust} considers the imperfect estimate of the target's direction and proposes a robust solution to the SRM problem. The simulation results are provided in Section~\ref{sec:num_results}. Finally, Section~\ref{sec:conclusions} concludes the paper.

Our notations are as follows. Upper (lower) bold face letters are used for matrices (vectors); $(\cdot)^T$ and $(\cdot)^H$ denote transpose and Hermitian transpose, respectively. $(\cdot)^*$ is the conjugate operator. $\mathbf{I}_N$ denotes the $N \times N$ identity matrix. For a complex-valued vector $\bm x$, $\| \bm x \|$ denotes the Euclidean norm. $\mathbb{H}^N$ and $\mathbb{R}$ denote the space of $N \times N$ Hermitian matrices and one-dimensional space of real numbers, respectively. ${\rm Tr}(\cdot)$ denotes a trace operation and $\mathbf{A}\succeq \mathbf{0}$ means that $\mathbf{A}$ is Hermitian positive semidefinite. The distribution of a circularly symmetric complex Gaussian (CSCG) random vector with mean vector $\bm x$ and covariance matrix $\bm \Sigma$ is denoted by $\mathcal{CN} ({\bm x}, {\bm \Sigma})$.

\section{System Model and Problem Formulation} \label{sec:model}

%Consider the multiuser downlink secnario, where an IRS is deployed to assist in the communications from a multi-antenna BS to $K$ single-antenna users. The number of transmit antennas at the BS and the reflecting units at the IRS are denoted by $N$ and $M$, respectively. Each of passive reflecting elements can be software-controlled to induce a desired phase shift on the incident signal.

\begin{figure}[h]
	\centering
	\centerline{\resizebox{.5\textwidth}{!}{\includegraphics{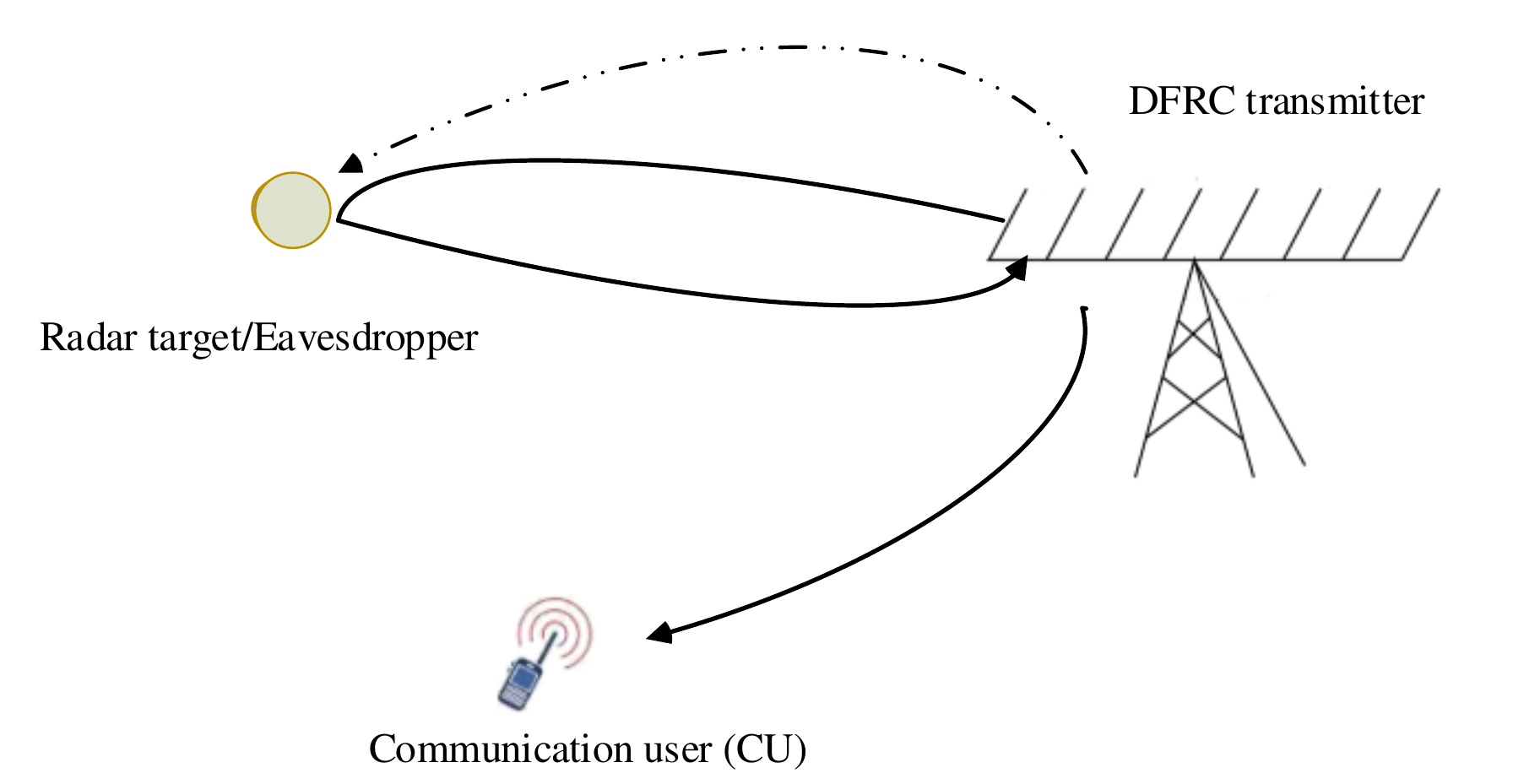}}} 
	\caption{Joint communication and radar system.}
	\label{scenarios}
\end{figure}

Consider a joint communication and radar system in Figure~\ref{scenarios}, which consists of a DFRC transmitter, a target and a CU. The DFRC aims to transmit information to the CU, and meanwhile exploit the information-embedded signal to estimate the distance of the target from the echo. We assume that the target is quasi-static within each slow time, and that the target could potentially be an eavesdropper, i.e., the target could also intercept the communication information from the received waveform. To prevent eavesdropping, the PLS technology is employed at the DFRC transmitter. Specifically, we assume that the DFRC transmitter has multiple antennas, and the CU and the target have a single antenna to receive information. The transmit signal $\bm x(t) \in \mathbb{C}^N$ at the DFRC is given by
\begin{equation} \label{eq:tx_sig}
\bm x(t) = \sqrt{P} \bm w s(t)
\end{equation} 
where $\bm  w \in \mathbb{C}^N$ is the transmit beamformer satisfying $\| \bm w \|=1$, $N$ is the number of transmit antennas at the DFRC transmitter, $s(t)$ with  ${\mathbb E}[|s(t)|^2]=1$ is the information encoded pulse waveform, and $P>0$ is the transmit power.

The received signal at the target is given by
\begin{equation} \label{eq:rx_sig_target}
y_1(t) = \beta_1 \bm h(\theta_1)^H \bm x(t) + n_1(t),
\end{equation}
where $\bm h(\theta) \triangleq [1, ~e^{\jmath\frac{2\pi}{\lambda}d\sin(\theta)},\ldots, e^{\jmath\frac{2\pi}{\lambda}(N-1)d\sin(\theta)}] \in \mathbb{C}^N$ denotes the steering vector of the transmit antenna array with $d$ being the inter-antenna spacing and $\lambda>0$ being the carrier wavelength; $\beta_1 \in \mathbb{C}$ is the complex path-loss coefficient; $\theta_1\in [-\pi/2, ~\pi/2]$ is the angle of departure from the DFRC transmitter to the target, and $n_1(t)\sim {\cal CN}(0,\sigma^2)$  is additive white Gaussian noise.

Similarly, the received signal at the CU is given by
\begin{equation} \label{eq:rx_sig_CR}
y_2(t) = \beta_2 \bm h(\theta_2)^H \bm x(t) + n_2(t),
\end{equation}
where $\beta_2\in \mathbb{C}$ is the complex path-loss coefficient from the DFRC transmitter to the CU, $\theta_2 \in [-\pi/2, ~\pi/2]$ is the angle of departure from the DFRC transmitter to the CU and $n_2(t) \sim {\cal CN}(0,\sigma^2) $  is additive white Gaussian noise at the CU.

%At the JCR, the returned echo signal $\bm y_3(t) \in \mathbb{C}^N$  is expressed as
%\begin{equation}
%\bm y_3(t) = \beta_3 \bm h(\theta_1) \bm h(\theta_1)^H \bm x(t-\tau) + \bm n_3(t),
%\end{equation} 
%where $\tau$ is the round-trip delay, $\beta_3 \in \mathbb{C}$ represents the complex combined gain (including antenna gain, cross-section, and propagation loss),  $\bm n_3(t) \sim {\cal CN}(\bm 0, \sigma^2 \bm I)$. 

At the DFRC, the received echo signal after receive beamforming is expressed as
\begin{equation}
y_3(t) =\bm f^H   (\beta_3  \bm h(\theta_1) \bm h(\theta_1)^H \bm x(t-\tau) + \bm n_3(t)),
\end{equation} 
where $\bm f \in \mathbb{C}^N$ is the receive beamformer, satisfying $\| \bm f \| = 1$; $\tau$ is the round-trip delay, $\beta_3 \in \mathbb{C}$ represents the combined  path-loss (including antenna gain, cross-section, and propagation loss),  $\bm n_3(t) \sim {\cal CN}(\bm 0, \sigma^2 \bm I)$. 

%\begin{figure}[h]
%	\centering
%	\centerline{\resizebox{.49\textwidth}{!}{\includegraphics{ISAC_model.pdf}}} 
%	\caption{Joint communication-radar system.}
%	\label{scenarios}
%\end{figure}

Traditionally, target detection  and  secure communication are studied in different context with different performance measures. For the DFRC system, we need a unified and tractable performance metric to describe the two tasks within one theme. To this end, inspired by~\cite{Bliss14}, we adopt the estimation rate and the secrecy rate as the performance measure of DFRC for secure communication and radar, respectively. Let us first give a brief introduction of the  two measures in the following.

\subsection{Estimation Rate} 
The concept of estimation rate was first introduced by Bliss in~\cite{Bliss14}. It characterizes the detection performance from an information-theoretic point of view. Specifically, according to the  estimation theory, the Cram\'er-Rao bound gives the minimum variance that an unbiased estimator can achieve. If the variance is regarded as the uncertainty of the parameter estimation, then similar to mutual information in communication, the ``elimination of target parameter uncertainty'' by radar can be defined as the mutual information  or the estimation rate (normalized by pulse repetition interval (PRI)) between the radar and the target. In a nutshell, the estimation rate reflects the target parameter entropy change before and after the estimation of that parameter;  readers are referred to~\cite{Bliss14,Bliss17} for a more detailed explanation. For multiple independent detection targets, each estimated target can be regarded as an independent information channel. Therefore, the radar estimation rate $R_{\rm est}$ (bit/pulse repetition interval) of multiple detection targets is defined as~\cite{Bliss14}:
\begin{equation} \label{eq:est_rate_def}
R_{\rm est} \leq  \sum_{m=1}^M \frac{h_{\tau_m,{\rm rr}}-h_{\tau_m, {\rm est}}}{T_{\rm pri}}
\end{equation}
where $M$ is the number of targets; $T_{\rm pri}$ is the pulse repetition interval; $\tau_m$ is the estimation parameter for the $m$th target; $h_{\tau_m, {\rm est}}$ is the entropy of the estimation (error)  and $h_{\tau_m,{\rm rr}}$ is the entropy of the parameter itself. 
We should mention that $h_{\tau_m,{\rm rr}}$ and $h_{\tau_m, {\rm est}}$ characterize the uncertainty of the  parameter $\tau_m$ before and after estimation, respectively.

For the delay estimation under circularly symmetric Gaussian noise, the $h_{\tau_m, {\rm est}}$ is given by~\cite{Bliss14}
\begin{equation}
h_{\tau_m, {\rm est}} = \log_2(\pi e \sigma_{\tau_m, {\rm est}}^2 )
\end{equation}
where $\sigma_{\tau_m, {\rm est}}^2$ is the variance of the estimate. Particularly, for an unbaised estimator the minimum variance is given by the Cram\'er-Rao bound:
\begin{equation} \label{eq:entropy_est}
\sigma_{\tau_m, {\rm est}}^2 = \frac{\sigma^2}{\gamma^2 T B^3 |\beta|^2 P}
\end{equation}
where $\beta \in \mathbb{C}$ is the combined gain, $B$ is the bandwidth, $\gamma>0$ is a scaling constant, depending on the shape of the radar waveform's power spectral density.  

For the $h_{\tau_m,{\rm rr}}$, assuming Gaussian process variation on the underlying estimation parameter within each PRI, the entropy of the parameter $\tau_m$ can be calculated as
\begin{equation}\label{eq:entropy_rr}
h_{\tau_m,{\rm rr}} = \log_2(\pi e (\sigma_{\tau_m, {\rm proc}}^2 + \sigma_{\tau_m, {\rm est}}^2 ) )
\end{equation}
where $\sigma_{\tau_m, {\rm proc}}^2$ is the variance of the Gaussian process. By substituting~\eqref{eq:entropy_est} and \eqref{eq:entropy_rr} into \eqref{eq:est_rate_def}, we have 
\begin{equation}
\begin{aligned}
R_{\rm est} & \leq   \frac{1}{T_{\rm pri}}\sum_{m=1}^M \log_2 \left(1 + \frac{\sigma^2_{\tau_m,{\rm proc}}}{\sigma^2_{\tau_m, {\rm est}}} \right) \\
& =  \frac{\delta}{T}\sum_{m=1}^M \log_2 \left(1 +   \frac{\sigma^2_{\tau_m,{\rm proc}} \gamma^2 T B^3 |\beta|^2 P}{\sigma^2} \right) 
\end{aligned}
\end{equation}
where $T = \delta T_{\rm pri}$ is the pulse duration and $\delta \in (0,~1)$ is the duty factor.

\subsection{Secrecy Rate}
Secrecy rate measures the maximum  transmission rate at which the confidential information can be securely sent from the transmitter to the legitimate receiver in the presence of eavesdropper. For the Gaussian wiretap channels, assuming Gaussian random modulation the secrecy rate is given by~\cite{Liang_book}
%\begin{equation}
%R_{\rm sec} = \left[  \log_2(1+ {\sf SNR}_{\rm rx}) - \max_{i=1,\ldots, M}\log_2 (1+{\sf SNR}_{{\rm eve},i}) \right]^+
%\end{equation}
%where $[\cdot]^+ = \max\{0, \cdot \}$;  ${\sf SNR}_{\rm rx}$ and ${\sf SNR}_{{\rm eve},i}$ denote the received SNR at the legitimate receiver and the $i$th eavesdropper, resp.; 
\begin{equation*}
R_{\rm sec} = \left[  B \log_2(1+ {\sf SNR}_{\rm cu}) - B\log_2 (1+{\sf SNR}_{{\rm eve}}) \right]^+
\end{equation*}
where $[\cdot]^+ = \max\{0, \cdot \}$;  ${\sf SNR}_{\rm cu}$ and ${\sf SNR}_{{\rm eve}}$ denote the received SNR at the CU and the eavesdropper, respectively. When specifying $R_{\rm sec}$ to the model in~\eqref{eq:tx_sig}-\eqref{eq:rx_sig_CR}, we have
\begin{equation}
R_{\rm sec} = \left[ B \log_2 \left(\frac{ \sigma^2 + P|\beta_2|^2 |\bm h(\theta_2)^H \bm w|^2}{\sigma^2 + P|\beta_1|^2 |\bm h(\theta_1)^H \bm w|^2} \right) \right]^+.
\end{equation}

\subsection{Problem Formulation}
We consider the following SRM problem:
\begin{equation} \label{eq:main}
\begin{aligned}
\max_{\bm w, \bm f}  & ~ R_{\rm sec} \\
{\rm s.t.} & ~ R_{\rm est} \geq \zeta, \quad \| \bm w\| = 1, \quad \| \bm f \| =1,
\end{aligned}
\end{equation}
where $\zeta>0$ specifies the minimum estimation rate. In words, we aim to maximize the secrecy rate for the CU and meanwhile guarantee  estimation rate above pre-specified threshold to  satisfy estimation performance for the target.

Notice that the receive beamformer $\bm f$ appears only in $R_{\rm est}$. It is easy to see that the optimal $\bm f^\star$ is the matched filter $\bm f^\star =  \bm h(\theta_{1}) / \sqrt{N}$.
By substituting $\bm f^\star$ into $R_{\rm est}$, problem~\eqref{eq:main} can be rewritten as
\begin{equation} \label{eq:main_eqv}
\begin{aligned}
\max_{\bm w}  & ~\frac{\bm w^H \bm A \bm w}{\bm w^H \bm B \bm w} \\
{\rm s.t.} & ~  \bm w^H \bm C \bm w   \geq 1, \\
& ~   \bm w^H \bm w = 1,
\end{aligned}
\end{equation}
where 
$\bm A = \sigma^2\bm I +  P|\beta_2|^2  \bm h(\theta_2)  \bm h(\theta_2)^H $, $\bm B= \sigma^2 \bm I +  P|\beta_1|^2  \bm h(\theta_1)  \bm h(\theta_1)^H $, $\bm C =\frac{N \sigma^2_{\tau, {\rm proc}} |\beta_3|^2 \gamma^2 B^3 T P}{\sigma^2 (2^{T\zeta/\delta} -1)}  \bm h(\theta_1)  \bm h(\theta_1)^H $. Problem~\eqref{eq:main_eqv} is a nonconvex fractional quadratic optimization problem. In the next section, we will exploit the problem structure and develop an optimal solution to it.

\section{A Closed-form Solution to Problem~\eqref{eq:main_eqv}} \label{sec:solution_no_AN}
In the first subsection, we  show that~\eqref{eq:main_eqv} can be optimally solved with 
SDR. In the second subsection, we will further exploit the problem structure to obtain a closed-form solution to problem~\eqref{eq:main_eqv}.

\subsection{An SDR Approach to Problem~\eqref{eq:main_eqv}}
We first employ the SDR approach to solve problem~\eqref{eq:main_eqv}. Let $\bm W = \bm w \bm w^H$ and drop the rank-one constraint on $\bm W$ to get the SDR of problem~\eqref{eq:main_eqv}:
%\begin{equation} \label{eq:main_sdr}
%\begin{aligned}
%\max_{\bm W}  & ~\frac{{\rm Tr}( \bm A \bm W)}{{\rm Tr}( \bm B \bm W)} \\
%{\rm s.t.} & ~ {\rm Tr}( \bm C \bm W)     \geq 1, \\
%& ~  {\rm Tr}(\bm W) = 1, \\
%& ~ \bm W \succeq \bm 0.
%\end{aligned}
%\end{equation}
\begin{equation} \label{eq:main_sdr}
\begin{aligned}
\max_{\bm W}  & ~\frac{{\rm Tr}( \bm A \bm W)}{{\rm Tr}( \bm B \bm W)} \\
{\rm s.t.} & ~ {\rm Tr}( \bm C \bm W)     \geq 1, \quad  {\rm Tr}(\bm W) = 1, \quad \bm W\succeq \bm 0.  
\end{aligned}
\end{equation}
Problem~\eqref{eq:main_sdr} is a fractional SDP, which can be further reformulated as a standard SDP by applying the Charnes-Cooper transformation~\cite{Charnes1962}. Specifically, by making a change of variables $\bm X = \kappa \bm W$ with $\kappa\geq 0$,  problem~\eqref{eq:main_sdr} can be equivalently written as
%\begin{equation} \label{eq:main_sdp}
%\begin{aligned}
%\max_{\bm X, \kappa}  & ~{\rm Tr}( \bm A \bm X)  \\
%{\rm s.t.} & ~ {\rm Tr}( \bm B \bm X) = 1, \\
%& ~ {\rm Tr}( \bm C \bm X)     \geq \kappa, \\
%& ~  {\rm Tr}(\bm X) = \kappa, \\
%& ~ \bm X \succeq \bm 0, \quad \kappa \geq 0
%\end{aligned}
%\end{equation}
\begin{equation} \label{eq:main_sdp}
\begin{aligned}
\max_{\bm X, \kappa}  & ~{\rm Tr}( \bm A \bm X)  \\
{\rm s.t.} & ~ {\rm Tr}( \bm B \bm X) = 1, \quad {\rm Tr}( \bm C \bm X)     \geq \kappa, \\
& ~ {\rm Tr}(\bm X) = \kappa, \quad \bm X \succeq \bm 0, \quad \kappa \geq 0.
\end{aligned}
\end{equation}
The equivalence between problems~\eqref{eq:main_sdr} and \eqref{eq:main_sdp}  can be established by using a similar proof in~\cite{QLI2011}. Denote $(\bm X^\star, \kappa^\star)$ as the optimal solution. The optimal $\bm W^\star$ for problem~\eqref{eq:main_sdr} can be obtained as $\bm W^\star = \bm X^\star / \kappa^\star$. In general, solving the SDR may yield a high-rank optimal $\bm W^\star$ and there is a relaxation gap between problems~\eqref{eq:main_sdr} and \eqref{eq:main}. However, for problem~\eqref{eq:main_sdr}, it can be shown that we must have a rank-one optimal $\bm W^\star$. 
\begin{Claim}\label{fact}
	Suppose that problem~\eqref{eq:main_sdp} is feasible. Then,  there must exist a rank-one $\bm W^\star$ for problem~\eqref{eq:main_sdp}.
\end{Claim} 
Claim~\ref{fact} can be easily shown by using the rank-reduction theorem~\cite{Huang2009} for SDP; herein we just give a sketched proof. Suppose $(\bm X^\star, \kappa^\star)$ is an optimal solution of problem~\eqref{eq:main_sdp} with ${\rm rank}(\bm X^\star)>1$. Then, by fixing $\kappa = \kappa^\star$ in problem~\eqref{eq:main_sdp} and considering the resultant SDP problem w.r.t. $\bm X$, one can see that there are in total three linear constraints w.r.t. $\bm X$. It follows from~\cite[Theorem 3.2]{Huang2009}, we can construct another optimal $\bm X$ for problem~\eqref{eq:main_sdp}, say $\hat{\bm X}$, such that ${\rm rank}(\hat{\bm X} ) \leq \sqrt{3}$. Clearly, the rank of $\hat{\bm X}$ can either be one or zero. Since $\hat{\bm X}=\bm 0$  is infeasible, we must have ${\rm rank}(\hat{\bm X})=1$. We should mention that while Claim~\ref{fact} says the existence of the rank-one $\bm W^\star$, such a rank-one $\bm W^\star$ can be efficiently constructed (with polynomial complexity) by following the proof procedure of the rank-reduction theorem; readers are referred to~\cite{Huang2009} for the detailed construction procedure.

Claim~\ref{fact} reveals that the nonconvex SRM problem~\eqref{eq:main_eqv} is actually hidden convex and thus optimally solvable. However, from the computation perspective, solving the SDP~\eqref{eq:main_sdp} is still costly with a numerical solver. Moreover,  it is generally hard to interpret the physical meaning of the solution returned by the solver. In view of that, we will develop a more efficient, closed-form solution to problem~\eqref{eq:main} by exploiting the problem structure in the next subsection.

\subsection{A Closed-form Solution to Problem~\eqref{eq:main_eqv}}
For ease of exposition, let us denote 
\begin{equation} \label{eq:h_def}
\begin{aligned} 
\bm h_1 & = \sqrt{\frac{P |\beta_1|^2}{\sigma^2}} \bm h(\theta_1), \quad \bm h_2   = \sqrt{\frac{P |\beta_2|^2}{\sigma^2}} \bm h(\theta_2), \\
 \alpha & =   \frac{|\beta_1|^2(2^{T\zeta/\delta} -1)}{N \sigma^2_{\tau, {\rm proc}} |\beta_3|^2 \gamma^2 B^3 T}
\end{aligned}
\end{equation}
and rewrite problem~\eqref{eq:main_eqv} as
\begin{equation}\label{eq:main_fqcqp}
\begin{aligned}
\max_{\bm w} ~ & \frac{1+\bm w^H \bm h_2 \bm h_2^H \bm w}{1+\bm w^H \bm h_1 \bm h_1^H \bm w } \\
{\rm s.t.}~ &  \bm w^H \bm h_1 \bm h_1^H \bm w \geq \alpha , \quad  \bm w^H  \bm w =1
\end{aligned}
\end{equation}
Let us first characterize the structure of the optimal $\bm w^\star$.
\begin{Claim}\label{ob:1}
	The optimal $\bm w^\star$ of problem~\eqref{eq:main_fqcqp} takes the following form:
	\begin{equation}\label{eq:opt_beamformer}
	\bm w^\star =  \lambda_1^\star \bm u_1 + \lambda_2^\star \bm u_2
	\end{equation}
	for some $\lambda_1^\star, \lambda_2^\star\in \mathbb{C}$, where $\bm u_1 = \bm h_1/\| \bm h_1\|$ and $\bm u_2 =( \bm I - \bm u_1\bm u_1^H) \bm h_2/\| ( \bm I - \bm u_1\bm u_1^H) \bm h_2\| $. 
\end{Claim}
\noindent{\it Proof.}~See Appendix~\ref{app:A}. \hfill $\blacksquare$

By substituting \eqref{eq:opt_beamformer} into \eqref{eq:main} and noticing $\bm u_1 \perp \bm u_2$, $\| \bm u_1\| =\| \bm u_2\|=1$,  problem~\eqref{eq:main_fqcqp} is simplified as 
\begin{subequations}\label{eq:main-2}
	\begin{align}
	\max_{\lambda_1, \lambda_2} ~ & \frac{1+|\lambda_1 \bm h_2^H \bm u_1 + \lambda_2 \bm h_2^H \bm u_2 |^2 }{1+  |\lambda_1|^2 | \bm h_1^H \bm u_1|^2 }  \label{eq:main-2-a} \\
	{\rm s.t.}~ &  |\lambda_1|^2 | \bm h_1^H \bm u_1|^2 \geq \alpha \label{eq:main-2-b} \\
	~& |\lambda_1|^2+|\lambda_2|^2=1 \label{eq:main-2-c}.
	\end{align}
\end{subequations} 
The numerator in~\eqref{eq:main-2-a} is expanded as
\begin{equation} \label{eq:numerator_upper_bound}
\begin{aligned}
& 1+ |\lambda_1|^2 |\bm h_2^H \bm u_1|^2 + |\lambda_2|^2 |\bm h_2^H \bm u_2 |^2 \\
& + 2\Re\{ \lambda_1 \lambda_2^*\bm h_2^H \bm u_1  \bm u_2^H \bm h_2\}  \\
\leq & 1+ |\lambda_1|^2 |\bm h_2^H \bm u_1|^2 + |\lambda_2|^2 |\bm h_2^H \bm u_2 |^2 \\
& + 2|\lambda_1| | \lambda_2| |\bm h_2^H \bm u_1|  |\bm u_2^H \bm h_2|
\end{aligned}
\end{equation}
where the equality holds if 
\begin{equation}\label{eq:upper_bound_condition}
\angle \lambda_1 = -\angle (\bm h_2^H \bm u_1)~{\rm and}~\angle \lambda_2 =- \angle (\bm h_2^H \bm u_2).
\end{equation}
Notice that for any given $\lambda_1$ and $\lambda_2$, we can always rotate the phases of $\lambda_1$ and $\lambda_2$  such that \eqref{eq:upper_bound_condition} holds without changing the denominator in \eqref{eq:main-2-a} and the constraints~\eqref{eq:main-2-b}-\eqref{eq:main-2-c}. Therefore, without loss of optimality, we can assume that the optimal $\lambda_1^\star$ and $\lambda_2^\star$ take the following form:
\begin{equation}\label{eq:opt_lambda}
\lambda_1^\star = {\lambda}_1 e^{ - \jmath  \angle (\bm h_2^H \bm u_1)}, \quad \lambda_2^\star = {\lambda}_2 e^{ - \jmath  \angle (\bm h_2^H \bm u_2)}
\end{equation}  
for some $\lambda_i\in \mathbb{R}, i=1,2$. We make a change of variables 
\begin{equation} \label{eq:mu_def}
\theta = |\lambda_1|^2, ~~\mu_1 = |\bm h_2^H \bm u_1|,~~ \mu_2=|\bm h_2^H \bm u_2|, ~~\mu_3=|\bm h_1^H \bm u_1|.
\end{equation} Using the upper bound in~\eqref{eq:numerator_upper_bound}, problem~\eqref{eq:main-2} is simplified as
\begin{subequations}\label{eq:1-dim-opt}
	\begin{align}
	\max_{\theta} ~ &  f(\theta) \triangleq \frac{ 1+  (\sqrt{\theta}\mu_1 +\sqrt{1-\theta}\mu_2)^2}{1+  \mu_3^2 \theta}  \label{eq:1-dim-opt-a} \\
	{\rm s.t.}~ &   \alpha/\mu_3^2 \leq \theta \leq 1 . \label{eq:1-dim-opt-b}
	\end{align}
\end{subequations} 

Since problem~\eqref{eq:1-dim-opt} is a one-dimensional problem with box constraint. The optimal $\theta$ attains either at the points with vanishing gradient or at the boundary, i.e., $\theta=1$ or $\theta=\alpha/\mu_3^2$. Let us first check the  points with vanishing gradient, i.e.,
%\begin{equation}
%\begin{aligned}
%& \nabla_\theta f(\theta) = 0\\
%\Longleftrightarrow~ & \frac{\kappa_1 \sqrt{\theta (1-\theta)} - \kappa_2 \theta + \kappa_3}{\sqrt{\theta (1-\theta)} (1+ \mu_3^2 \theta)^2} = 0 \\
%\Longrightarrow~ & \theta_1 = \frac{(\kappa_1^2+2\kappa_2\kappa_3)+\sqrt{\kappa_1^4+ 4\kappa_1^2\kappa_2\kappa_3-4\kappa_1^2\kappa_3^2}}{2(\kappa_1^2+\kappa_2^2)} \\
%&  ~{\rm and}~\\
%&  \theta_2 = \frac{(\kappa_1^2+2\kappa_2\kappa_3)-\sqrt{\kappa_1^4+ 4\kappa_1^2\kappa_2\kappa_3-4\kappa_1^2\kappa_3^2}}{2(\kappa_1^2+\kappa_2^2)}
%\end{aligned}
%\end{equation}  
\begin{equation}
\begin{aligned}
 &\nabla_\theta f(\theta) = 0 \Longleftrightarrow \frac{\kappa_1 \sqrt{\theta (1-\theta)} - \kappa_2 \theta + \kappa_3}{\sqrt{\theta (1-\theta)} (1+ \mu_3^2 \theta)^2} = 0 \\
&\Longrightarrow \theta_1 = \frac{(\kappa_1^2+2\kappa_2\kappa_3)+\sqrt{\kappa_1^4+ 4\kappa_1^2\kappa_2\kappa_3-4\kappa_1^2\kappa_3^2}}{2(\kappa_1^2+\kappa_2^2)},\\
& \theta_2 = \frac{(\kappa_1^2+2\kappa_2\kappa_3)-\sqrt{\kappa_1^4+ 4\kappa_1^2\kappa_2\kappa_3-4\kappa_1^2\kappa_3^2}}{2(\kappa_1^2+\kappa_2^2)}
\end{aligned}
\end{equation}   
where 
$\kappa_1= \mu_1^2-(\mu_3^2+1)\mu_2^2 - \mu_3^2, \quad \kappa_2= \mu_1\mu_2(\mu_3^2+2), \quad \kappa_3 = \mu_1 \mu_2 $. Since $\theta_1+\theta_2 = \frac{\kappa_1^2+2\kappa_2\kappa_3}{\kappa_1^2+\kappa_2^2} \geq 0$ and $\theta_1 \theta_2 = \frac{\kappa_3^2}{\kappa_1^2+\kappa_2^2} \geq 0$, we have $\theta_1 \geq 0$ and $\theta_2 \geq 0$. Moreover, since $\kappa_2 \geq \kappa_3$, $\theta_1+\theta_2 = \frac{\kappa_1^2+2\kappa_2\kappa_3}{\kappa_1^2+\kappa_2^2} \leq 1$, which implies
$ 0 \leq \theta_2 \leq \theta_1 \leq 1 $. Therefore, the optimal $\theta^\star$ is given by
\[   \theta^\star  = {\arg\max}_{\theta \in \{ \alpha/\mu_3^2, 1, \theta_1, \theta_2 \} \cap  \{ \alpha/\mu_3^2\leq  \theta \leq 1 \} }  f(\theta). \]
With the optimal $\theta^\star$, the optimal $\bm w^\star$ can be obtained as
\begin{equation} \label{eq:opt_beamer_expression}
\bm w^\star  =  \sqrt{\theta^\star} e^{ - \jmath  \angle (\bm h_2^H \bm u_1)} \bm u_1 +  \sqrt{1-\theta^\star} e^{ - \jmath  \angle (\bm h_2^H \bm u_2)} \bm u_2.
\end{equation}

\section{Artificial Noise-aided DFRC Beamforming} \label{sec:solution_with_AN}
In this section, we consider incorporating AN into the transmission. With AN, the DFRC transmit signal~\eqref{eq:tx_sig} is modified as
\begin{equation}
\bm x(t) = \sqrt{p_1} \bm w s(t) + \sqrt{p_2} \bm  z(t)
\end{equation}
where $p_1+p_2 = P$, $\bm z(t)\in \mathbb{C}^{N}$  is the random distortional
waveform following complex  Gaussian distribution with mean zero and covariance $\bm \Phi \succeq \bm 0$ and ${\rm Tr}(\bm \Phi)  =1$.

Following a similar derivation  in the last section, the corresponding estimation rate  is given by~\eqref{eq:est_rate},
\begin{figure*}[!t]
	\setlength\arraycolsep{0pt}
	\begin{equation}\label{eq:est_rate}
	\begin{aligned}
	& R_{\rm est} =\frac{\delta}{T}\log_2 \left( 1+  \frac{\sigma^2_{\tau,{\rm proc}} \gamma^2 T B^3 N|\beta_3|^2 (p_1|\bm h(\theta_{1})^H \bm w|^2 + p_2 \bm h(\theta_{1})^H \bm \Phi \bm h(\theta_{1}))  }{\sigma^2}   \right)
	\end{aligned}
	\end{equation}
	\hrulefill
\end{figure*}
%\begin{equation}
%\begin{aligned}
%& R_{\rm est} =
%&  \frac{\delta}{T}\log_2 \left( 1+  \varpi \right)
%\end{aligned}
%\end{equation}
%where $\varpi = \frac{\sigma^2_{\tau,{\rm proc}} \gamma^2 T B^3 N|\beta_3|^2 (p_1|\bm h(\theta_{1})^H \bm w|^2 + p_2 \bm h(\theta_{1})^H \bm \Phi \bm h(\theta_{1}))  }{\sigma^2}$,
and the secrecy rate is given by
\begin{equation}
R_{\rm sec} =[ B \log_2 (1+ {\sf SINR}_{\rm cu}) -  B \log_2(1+ {\sf SINR}_{\rm eve}) ]^+,
\end{equation}
where
\begin{subequations} \label{eq:SINR}
	\begin{align}
	{\sf SINR}_{\rm cu}= &   \frac{ p_1 |\beta_2|^2 | \bm h(\theta_{2})^H \bm w |^2  }{\sigma^2+ p_2 |\beta_2|^2  \bm h(\theta_{2})^H \bm \Phi  \bm h(\theta_{2}) },  \label{eq:SINR_a}  \\
	{\sf SINR}_{\rm eve} = &   \frac{p_1 |\beta_1|^2 | \bm h(\theta_{1})^H \bm w |^2  }{\sigma^2+ p_2 |\beta_1|^2  \bm h(\theta_{1})^H \bm \Phi \bm h(\theta_{1})  }. \label{eq:SINR_b}
	\end{align}
\end{subequations}

%\begin{equation}
%\begin{aligned}
%& R_{\rm sec} =\\
%& \left[ B \log_2 \left( 1+  \frac{ p_1 |\beta_2|^2 | \bm h(\theta_{2})^H \bm w |^2  }{k_{\rm B} T_{\rm temp} B+ p_2 |\beta_2|^2 | \bm h(\theta_{2})^H \bm \varphi |^2  }   \right)  \right. \\
%& \left. -   B \log_2 \left( 1+  \frac{p_1 |\beta_1|^2 | \bm h(\theta_{1})^H \bm w |^2  }{k_{\rm B} T_{\rm temp}B+ p_2 |\beta_1|^2 | \bm h(\theta_{1})^H \bm \varphi |^2  }   \right)  \right]^+ .
%\end{aligned}
%\end{equation}

The AN-aided SRM problem~\eqref{eq:main_eqv} is formulated as
\begin{subequations}\label{eq:main_AN}
	\begin{align}
	\max_{p_1, p_2, \bm w, \bm \Phi} & ~ g(p_1, p_2, \bm w, \bm \Phi) \label{eq:main_AN_a} \\
	{\rm s.t.} &~ p_1|\bm h(\theta_{1})^H \bm w|^2 + p_2 \bm h(\theta_{1})^H \bm \Phi \bm h(\theta_{1}) \notag \\
	& ~  \geq  \bar{\zeta} , \label{eq:main_AN_b}\\
	& ~ p_1 + p_2 = P, \label{eq:main_AN_c}\\
	&  \| \bm w \| = 1, \quad  {\rm Tr}(\bm \Phi)  = 1, \label{eq:main_AN_d}\\
	& p_1 \geq  0, \quad p_2 \geq 0, \quad \bm \Phi \succeq \bm 0, \label{eq:main_AN_e}
	\end{align}
\end{subequations}
%where $g( p_1, p_2, \bm w, \bm \Phi) = \frac{g_1( p_1, p_2, \bm w, \bm \Phi)}{g_2( p_1, p_2, \bm w, \bm \Phi)}$ and $g_1( p_1, p_2, \bm w, \bm \Phi) =( \sigma^2+ p_2 |\beta_2|^2 \bm h(\theta_{2})^H \bm \Phi \bm h(\theta_{2})   +  p_1 |\beta_2|^2 | \bm h(\theta_{2})^H \bm w |^2) ( \sigma^2 + p_2 |\beta_1|^2  \bm h(\theta_{1})^H \bm \Phi \bm h(\theta_{1})) $ and $g_2( p_1, p_2, \bm w, \bm \Phi) =  (\sigma^2 + p_2 |\beta_2|^2  \bm h(\theta_{2})^H \bm \Phi \bm h(\theta_{2})) (\sigma^2 + p_2 |\beta_1|^2 \bm h(\theta_{1})^H \bm \Phi \bm h(\theta_{1}) +  p_1 |\beta_1|^2 | \bm h(\theta_{1})^H \bm w |^2)$, and $\bar{\zeta} = \frac{\sigma^2( 2^{T\zeta/\delta}-1)}{\sigma^2_{\tau,{\rm proc}} \gamma^2 T B^3 N|\beta_3|^2}$.
where $\bar{\zeta} = \frac{\sigma^2( 2^{T\zeta/\delta}-1)}{\sigma^2_{\tau,{\rm proc}} \gamma^2 T B^3 N|\beta_3|^2}$, $g( p_1, p_2, \bm w, \bm \Phi) = \frac{g_1( p_1, p_2, \bm w, \bm \Phi)}{g_2( p_1, p_2, \bm w, \bm \Phi)}$ and 
\begin{align*}
&g_1( p_1, p_2, \bm w, \bm \Phi)  =( \sigma^2+ p_2 |\beta_2|^2 \bm h(\theta_{2})^H \bm \Phi \bm h(\theta_{2})   + \\
& ~ p_1 |\beta_2|^2 | \bm h(\theta_{2})^H \bm w |^2) ( \sigma^2 + p_2 |\beta_1|^2  \bm h(\theta_{1})^H \bm \Phi \bm h(\theta_{1})),  \\
&g_2( p_1, p_2, \bm w, \bm \Phi)  =  (\sigma^2 + p_2 |\beta_2|^2  \bm h(\theta_{2})^H \bm \Phi \bm h(\theta_{2})) \times  \\
& ~ (\sigma^2 + p_2 |\beta_1|^2 \bm h(\theta_{1})^H \bm \Phi \bm h(\theta_{1}) +  p_1 |\beta_1|^2 | \bm h(\theta_{1})^H \bm w |^2).
\end{align*}

Let us first analyze the structure of the AN's covariance matrix $\bm \Phi$. We have the following claim.
%Notice that $\bm \varphi$ affects problem~\eqref{eq:main_AN} through $\bm h^H(\theta_{1}) \bm \varphi$, $\bm h^H(\theta_{2}) \bm \varphi$. Therefore, without loss of optimality, the optimal $\bm \varphi$ can be expressed as
%\[ \bm \varphi^\star =  \eta_1^\star \bm u_1 + \eta_2^\star \bm u_2\]
%where $\bm u_1$ and $\bm u_2$ are defined in~\eqref{eq:opt_beamformer}.  Since $\bm u_2$  is orthogonal to $\bm u_1$ (or $\bm h(\theta_{1})$), $\eta_2\bm u_2$ affects only the legitimate communication user's SINR in an adversary manner. Clearly, at the optimal $\bm \varphi^\star$, we should set $\eta_2\bm u_2 = \bm 0$. Therefore, we have the following observation.
\begin{Claim}\label{ob:2}
	Let $\bm \Phi^\star = \bm u_1 \bm u_1^H$. Then, $\bm \Phi^\star$ is optimal for problem~\eqref{eq:main_AN}, i.e., the optimal AN can be generated by transmit beamforming. 
\end{Claim}
\noindent{\it Proof.}~See Appendix~\ref{app:ob2}. \hfill $\blacksquare$

Upon Claim~\ref{ob:2}, problem~\eqref{eq:main_AN} is simplified as
\begin{subequations}\label{eq:main_AN_simp}
	\begin{align}
	\max_{p_1, p_2, \bm w} & ~ \frac{(P+p_2 \mu_3^2)(P+\mu_1^2 p_2 + p_1 |\bm h_2^H\bm w|^2)}{(P+p_2 \mu_1^2)(P+\mu_3^2 p_2 + p_1 |\bm h_1^H\bm w|^2)} \label{eq:main_AN_simp_a}\\
	{\rm s.t.} &~ p_1|\bm h(\theta_{1})^H \bm w|^2 + N p_2 \geq  \bar{\zeta} , \label{eq:main_AN_simp_b} \\
	& ~ p_1 + p_2 = P, \label{eq:main_AN_simp_c}\\
	&  \| \bm w \| = 1, \quad p_1 \geq  0, \quad p_2 \geq 0,\label{eq:main_AN_simp_d}
	\end{align}
\end{subequations}
where $\bm h_1$, $\bm h_2$, $\mu_1$ and $\mu_3$ are defined in~\eqref{eq:h_def} and \eqref{eq:mu_def}, respectively. Problem~\eqref{eq:main_AN_simp} is still difficult to deal with.  Below we will develop an alternating optimization algorithm to iteratively update $p_1$, $p_2$ and $\bm w$. 

\subsection{Optimizing $\bm w$ for fixed $(p_1, p_2)$}
For fixed $p_1$ (also $p_2$), optimizing problem~\eqref{eq:main_AN_simp} w.r.t. $\bm w$ amounts to
\begin{equation}\label{eq:main_AN_simp_w}
\begin{aligned}
\max_{\bm w} & ~ \frac{ 1 +  \bm w^H \tilde{\bm h}_2 \tilde{\bm h}_2^H \bm w }{ 1 +  \bm w^H \tilde{\bm h}_1 \tilde{\bm h}_1^H \bm w } \\
{\rm s.t.} &~ \bm w^H \tilde{\bm h}_1 \tilde{\bm h}_1^H \bm w \geq  \tilde{\alpha}, \quad \bm w^H \bm w = 1,
\end{aligned}
\end{equation}
where
%\begin{align*}
%\tilde{\bm h}_1 & = \sqrt{p_1/(P+\mu_3^2 p_2)} \bm h_1 \\
%\tilde{\bm h}_2 & = \sqrt{p_1/(P+\mu_1^2 p_2)} \bm h_2 \\
%\tilde{\alpha} & = \frac{P  |\beta_1|^2 (\bar{\zeta} - N p_2)  }{\sigma^2 (P+ \mu_3^2 p_2)}.
%\end{align*}
$\tilde{\bm h}_1  = \sqrt{p_1/(P+\mu_3^2 p_2)} \bm h_1$, $\tilde{\bm h}_2 = \sqrt{p_1/(P+\mu_1^2 p_2)} \bm h_2$, $\tilde{\alpha}= \frac{P  |\beta_1|^2 (\bar{\zeta} - N p_2)  }{\sigma^2 (P+ \mu_3^2 p_2)}$.
Clearly, problem~\eqref{eq:main_AN_simp_w} has the same form as problem~\eqref{eq:main_fqcqp}, which can be optimally solved with the method developed in the last section; we skip the details for brevity.

\subsection{Optimizing $(p_1, p_2)$ for fixed $\bm w$}
Next, we consider optimizing the power allocation for fixed $\bm w$ in problem~\eqref{eq:main_AN_simp}, viz.,
\begin{equation}\label{eq:main_AN_simp_Power}
\begin{aligned}
\max_{p_1, p_2} & ~ \frac{(P+p_2 \mu_3^2)(P+\mu_1^2 p_2 + p_1 \psi_2^2)}{(P+p_2 \mu_1^2)(P+\mu_3^2 p_2 + p_1 \psi_1^2)} \\
{\rm s.t.} &~ p_1\psi_3^2 + N p_2 \geq  \bar{\zeta} , \\
& ~ p_1 + p_2 = P, \quad p_1 \geq  0, \quad p_2 \geq 0, 
\end{aligned}
\end{equation}
where $\psi_1 =  |\bm h_1^H\bm w|$, $\psi_2 = |\bm h_2^H\bm w|$ and $\psi_3 =|\bm h(\theta_{1})^H \bm w|$. Make use of \eqref{eq:main_AN_simp_c}  to eliminate $p_1$ and simplify problem~\eqref{eq:main_AN_simp_Power} as
\begin{equation}\label{eq:main_Power}
\begin{aligned}
\max_{p_2} & ~  g(p_2)  \triangleq \frac{a_1x^2 +b_1x +c_1}{a_2x^2 +b_2x +c_2}\\
{\rm s.t.} &~  \hat{\zeta}  \leq p_2 \leq P, 
\end{aligned}
\end{equation}
where $\hat{\zeta} = \max\{ 0, ~(\bar{\zeta}-P\psi_3^2)/(N-\psi_3^2)\}$, $a_1 = \mu_3^2(\mu_1^2-\psi_2^2)$, $b_1 = P(\mu_1^2 -\psi_2^2+\mu_3^2 + \mu_3^2\psi_2^2)$, $c_1 = (1+\psi_2^2)P^2$, $a_2 =\mu_1^2(\mu_3^2-\psi_1^2) $, $b_2 =P(\mu_3^2 -\psi_1^2+\mu_1^2 + \mu_1^2\psi_1^2)$ and $c_2 = (1+\psi_1^2)P^2$. 
%\[g(p_2) =  \frac{\mu_3^2(\mu_1^2-\psi_2^2)p_2^2 + P(\mu_1^2 -\psi_2^2+\mu_3^2 + \mu_3^2\psi_2^2)p_2+ (1+\psi_2^2)P^2}{\mu_1^2(\mu_3^2-\psi_1^2)p_2^2 + P(\mu_3^2 -\psi_1^2+\mu_1^2 + \mu_1^2\psi_1^2)p_2+ (1+\psi_1^2)P^2}. \] 
Problem~\eqref{eq:main_Power} is a fractional quadratic program  with a box constraint. Similar to problem~\eqref{eq:1-dim-opt}, the optimal $p_2^\star$ should be attained either  at the points  with vanishing gradient, i.e., $\nabla_{p_2} g(p_2^\star) = 0$ or at the boundary, i.e., $ p_2^\star = \hat{\zeta}$ or $p_2^\star =1$. Let us calculate the points with vanishing gradient: 
$\nabla_{p_2} g(p_2) = 0$, and the solution is given by~\eqref{eq:solution_p2}.
%in~\eqref{eq:solution_p2}.
\begin{figure*}[!t]
\setlength\arraycolsep{0pt}
\begin{subequations} \label{eq:solution_p2}
	\begin{align}
	p_2^+ & =   \frac{a_2 c_1 - a_1 c_2 + \sqrt{a_1^2 c_2^2 - 2 a_1 a_2 c_1 c_2 -a_1 b_1 b_2 c_2 + a_1 b_2^2 c_1 + a_2^2 c_1^2 + a_2 b_1^2 c_2 -a_2 b_1 b_2 c_1 }}{a_1 b_2 - a_2 b_1}, \label{eq:solution_p2_a} \\
	p_2^- & =   \frac{a_2 c_1 - a_1 c_2 - \sqrt{a_1^2 c_2^2 - 2 a_1 a_2 c_1 c_2 -a_1 b_1 b_2 c_2 + a_1 b_2^2 c_1 + a_2^2 c_1^2 + a_2 b_1^2 c_2 -a_2 b_1 b_2 c_1 }}{a_1 b_2 - a_2 b_1}.
	\label{eq:solution_p2_b} 
	\end{align}
\end{subequations}
\hrulefill
\end{figure*}
Therefore, the optimal $p_2^\star$ is given by:
\begin{equation} \label{eq:opt_power_solution}
p_2^\star = {\arg\max}_{p_2 \in \{ \hat{\zeta}, P, p_2^-, p_2^+  \} \cap \{ \hat{\zeta} \leq  p_2 \leq P  \} } g(p_2)
\end{equation} 
and the optimal $p_1^\star = P- p_2^\star$.

Algorithm~\ref{AL1} summarizes the AO procedure for problem~\eqref{eq:main_AN}. Since each AO update yields a non-decreasing objective value, and moreover the objective value in~\eqref{eq:main_AN_a}  is upper bounded due to the boundedness of the feasible set of problem~\eqref{eq:main_AN} , it follows from the  Bolzano–Weierstrass theorem that the AO Algorithm 1 is convergent.

\begin{algorithm}[H]
	\caption{AO Algorithm for Problem~\eqref{eq:main_AN}} 
	\begin{algorithmic}[1]\label{AL1}
		\STATE  {\bf Input}: an initial power allocation $(p_1^0, p_2^0)$,  iteration index $\ell=0$, threshold $\epsilon>0$;
		\REPEAT
		\STATE  Compute $\bm w^{\ell+1}$ from Eqn.~\eqref{eq:opt_beamer_expression} for $(p_1, p_2) = (p_1^\ell, p_2^\ell)$;
		\STATE  Compute $(p_1^{\ell+1}, p_2^{\ell+1})$ from Eqn.~\eqref{eq:opt_power_solution} for $\bm w  = \bm w^{\ell+1}$;
		\STATE $\ell = \ell+1$;		
		\UNTIL $|g(p_1^{\ell}, p_2^{\ell}, \bm w^{\ell}, \bm \Phi^\star) -g(p_1^{\ell-1}, p_2^{\ell-1}, \bm w^{\ell-1}, \bm \Phi^\star)|<\epsilon $
		\STATE  {\bf Output} $p_1^{\ell}, p_2^{\ell}, \bm w^{\ell}$
	\end{algorithmic}
\end{algorithm}

\section{Robust DFRC Transmit Beamforming}  \label{sec:robust}
In previous sections, we have assumed that the direction $\theta_{1}$ of the target is known exactly. In practice, this may not be possible due to estimation error. In this section we consider robust AN-aided DFRC beamforming by taking into account the direction error on the target. Specifically, we consider the following phase error model~\cite{So16,Ottersten15}:
\begin{equation}\label{eq:phase_error_model}
\bm h(\theta_1)  = \bm h(\bar{\theta}_1) \odot  e^{\jmath  \Delta \bm\theta } 
\end{equation}
where $\odot$ is the entry-wise product, $\bar{\theta}_1$ is the estimate of the target's direction and $e^{\jmath  \Delta \bm\theta}=  [e^{\jmath \Delta \theta_0}, \ldots, e^{\jmath \Delta \theta_{N-1}}]$. Here, $\Delta \bm \theta$ is  random phase error, following $\Delta \bm \theta \sim  {\cal P} \in  \mathscr{D}(\bar{\bm \mu}, \bm \Sigma)$ where  $\mathscr{D}(\bar{\bm \mu}, \bm \Sigma)$ denotes a probability distribution set with mean $\bar{\bm \mu}$ and covariance $\bm \Sigma$. That is, we know only the mean $\bar{\bm \mu}$ and covariance matrix $\bm \Sigma$ but the exact distribution ${\cal P}$ is unknown.

Upon~\eqref{eq:phase_error_model}, the robust AN-aided DFRC beamforming problem is formulated as follows.
\begin{subequations} \label{eq:main_robust}
	\begin{align}
	\max_{\bm w, \bm \Phi, \xi}  & ~ \xi \\
	{\rm s.t.} & ~ \inf_{{\cal P} \in  \mathscr{D}(\bar{\bm \mu}, \bm \Sigma) }{\rm Pr}_{\Delta \bm \theta \sim {\cal P}}\{ R_{\rm sec} \geq \xi \} \geq 1- \epsilon_{\rm sec} , \label{eq:main_robust_b}\\
	& ~ \inf_{{\cal P} \in  \mathscr{D}(\bar{\bm \mu}, \bm \Sigma) }{\rm Pr}_{\Delta \bm \theta \sim {\cal P}}\{ R_{\rm est} \geq \zeta \} \geq 1- \epsilon_{\rm est}, \label{eq:main_robust_c} \\
	& ~ \| \bm w \|^2 + {\rm Tr}(\bm \Phi) \leq P,
	\end{align}
\end{subequations}
where $0<\epsilon_{\rm sec}<0.5$ and $0<\epsilon_{\rm est}<0.5$ specify the outage probabilities (evaluated w.r.t. $\bm \Delta \bm \theta$) of secrecy rate and estimation rate, respectively. In words, the robust DFRC beamforming problem~\eqref{eq:main_robust} aims at maximizing the secrecy rate (with outage probability less than $\epsilon_{\rm sec}$), and meanwhile  guarantee estimation rate no smaller than $\zeta$ with probability at least $1-\epsilon_{\rm est}$. 

To begin with, let us recast problem~\eqref{eq:main_robust} into a more amenable form. Since $\bm \Delta \bm \theta$ appears only in $\bm h(\theta_1)$, problem~\eqref{eq:main_robust} can be rewritten as
\begin{subequations} \label{eq:main_robust_eqv}
	\begin{align}
	\max_{\bm w, \bm \Phi, \eta}  & ~ B\log_2(1+ {\sf SINR}_{\rm cu})  - B \log_2(1+ \eta ) \label{eq:main_robust_eqv_a} \\
	{\rm s.t.} & ~ \inf_{{\cal P} \in  \mathscr{D}(\bar{\bm \mu}, \bm \Sigma) }{\rm Pr}_{\Delta \bm \theta \sim {\cal P}}\{ {\sf SINR}_{\rm eve}  \leq \eta \} \geq 1- \epsilon_{\rm sec} , \label{eq:main_robust_eqv_b}\\
	& ~ \inf_{{\cal P} \in  \mathscr{D}(\bar{\bm \mu}, \bm \Sigma) }{\rm Pr}_{\Delta \bm \theta \sim {\cal P}}\{ R_{\rm est} \geq \zeta \} \geq 1- \epsilon_{\rm est},  \label{eq:main_robust_eqv_c} \\
	& ~ \| \bm w \|^2 + {\rm Tr}( \bm \Phi ) \leq P, \label{eq:main_robust_eqv_d}
	\end{align}
\end{subequations}
Let $\bm W = \bm w \bm w^H$. Notice that 
\begin{equation}\label{eq:robust_obj_eqv}
\begin{aligned}
& \eqref{eq:main_robust_eqv_a} \Longleftrightarrow  \max  \frac{1+ {\sf SINR}_{\rm cu}}{1+ \eta} \\
& \Longleftrightarrow   \max \frac{\sigma^2 + |\beta_2|^2 \bm h(\theta_2)^H (\bm W   + \bm \Phi) \bm h(\theta_2) }{(1+\eta)(\sigma^2 +  |\beta_2|^2 \bm h(\theta_2)^H \bm \Phi \bm h(\theta_2) )}.
\end{aligned}   
\end{equation}

\begin{equation}\label{eq:robust_sec_eqv}
\begin{aligned}
&  \eqref{eq:main_robust_eqv_b} \\
 =&  {\rm Pr}_{\Delta \bm \theta \sim {\cal P}} \left\{ \bm h(\theta_{1})^H (\bm W - \eta \bm \Phi) \bm h(\theta_{1}) \leq \frac{\eta \sigma^2 }{|\beta_1|^2} \right\}\\
= &  {\rm Pr}_{\Delta \bm \theta \sim {\cal P}} \left\{ (e^{\jmath \Delta \bm \theta})^H  {\cal M}(\bm W - \eta \bm \Phi) (e^{\jmath \Delta \bm \theta}) \leq \frac{\eta \sigma^2 }{|\beta_1|^2}    \right\} \\
\geq & 1-\epsilon_{\rm sec}
\end{aligned}   
\end{equation}
where $ {\cal M}(\cdot)$ is the operator on the space of Hermitian matrices given by
\[  {\cal M}(\bm X) = \bm X  \odot (\bm h(\bar{\theta}_1) \bm h(\bar{\theta}_1)^H)^T \]
and
\begin{equation}\label{eq:robust_est_eqv}
\begin{aligned}
\eqref{eq:main_robust_eqv_c}= & {\rm Pr}_{\Delta \bm \theta \sim {\cal P}} \left\{ \bm h(\theta_{1})^H (\bm W + \bm \Phi) \bm h(\theta_{1}) \geq \bar{\zeta} \right\}  \\
= &    {\rm Pr}_{\Delta \bm \theta \sim {\cal P}} \left\{   (e^{\jmath \Delta \bm \theta})^H  {\cal M}(\bm W+\bm \Phi) (e^{\jmath \Delta \bm \theta})  \geq \bar{\zeta}  \right\}  \\
\geq &   1-\epsilon_{\rm est}.
\end{aligned}   
\end{equation}

The outage probability constraints in \eqref{eq:robust_sec_eqv} and \eqref{eq:robust_est_eqv} generally cannot be expressed in an explicit form. In the following, we take a safe approximation approach to convert the  outage probability constraints into a more tractable form. We need the  following lemma:
\begin{Lemma}[\cite{So16}] \label{lemma:1}
	Let $\bm \theta\in \mathbb{R}^N, e^{\jmath \bm \theta} = [e^{\jmath \theta_1}, \ldots, e^{\jmath \theta_N}]^T$, and $\bm X = \bm X^{\Re} + \jmath \cdot \bm X^\Im$ be an $N\times N$ Hermitian matrix with $\bm X^{\Re} $ and $\bm X^\Im$ being the real and the imaginary part, respectively. Then, the second-order Taylor approximation of $(e^{\jmath \bm \theta})^H \bm X (e^{\jmath \bm \theta})$ is given by
	\begin{equation}\label{eq:lemma}
	(e^{\jmath \bm \theta})^H \bm X (e^{\jmath \bm \theta}) \approx \sum_{k,\ell} X_{k\ell} + \bm \theta^T {\cal L}(\bm X^\Re) \bm \theta + {\cal F}(\bm X^\Im)^T \bm \theta
	\end{equation} 
	where ${\cal L}: \mathbb{R}^{N\times N} \rightarrow \mathbb{R}^{N\times N}$ denotes a linear mapping given by
	\[ [{\cal L}(\bm A)]_{k \ell} = \begin{cases}
	A_{kk} - \sum_{j} A_{kj} & \mbox{for}~k=\ell , \\
	A_{k\ell} & \mbox{for}~k \neq \ell.
	\end{cases}  \]
	and ${\cal F}: \mathbb{R}^{N\times N} \rightarrow \mathbb{R}^N$ denotes a linear mapping given by
	\[ [{\cal F}(\bm B)]_k = 2 \sum_{j} B_{kj}. \]
\end{Lemma}
Now, by using Lemma~\ref{lemma:1} we can approximate the outage probability constraint \eqref{eq:robust_sec_eqv} as
\begin{equation}\label{eq:robust_sec_approx}
\begin{aligned}
& {\rm Pr} \left\{ (e^{\jmath \Delta \bm \theta})^H  {\cal M}(\bm W - \eta \bm \Phi) (e^{\jmath \Delta \bm \theta}) \leq {\eta \sigma^2 }/{|\beta_1|^2}    \right\} \\
\approx & {\rm Pr} \left\{   \Delta \bm \theta^T {\cal L}( {\cal M}^{\Re}(\bm W - \eta \bm \Phi) )  \Delta \bm \theta  + \right.\\
& \sum_{k,\ell}[ {\cal M}(\bm W - \eta \bm \Phi)]_{k\ell}  +  {\cal F}(  {\cal M}^{\Im}(\bm W - \eta \bm \Phi)  )^T   \Delta \bm \theta \\
&  \left.  \leq {\eta \sigma^2 }/{|\beta_1|^2} \right\} \\
 \geq &    1-\epsilon_{\rm sec},
\end{aligned}   
\end{equation}
where $ {\cal M}^{\Re}(\cdot)$ and  ${\cal M}^{\Im}(\cdot)$ denote the real and the imaginary part of ${\cal M}(\cdot)$, respectively. The inequality inside the bracket of \eqref{eq:robust_sec_approx} is quadratic w.r.t. the random vector $\Delta \bm \theta$. To turn~\eqref{eq:robust_sec_approx} into a more tractable form, we introduce the following theorem. 
\begin{Theorem} \label{lemma:drb}
	Let $\bm x\in \mathbb{C}^N$ be a random vector, following $\bm x \sim {\cal  P} \in \mathscr{D}(\bar{\bm \mu}, \bm \Sigma)$. Consider the quadratic function $f(\bm x) = \bm x^H \bm A \bm x + 2\Re(\bm b^H \bm x) + c$, where $\bm A \in \mathbb{H}^N$, $\bm b \in \mathbb{C}^N$ and $c \in \mathbb{R}$.  Then, 
	\begin{equation}\label{eq:lemma_drb_original}
	\inf_{{\cal P} \in \mathscr{D}(\bar{\bm \mu}, \bm \Sigma)}{\rm Pr}_{\bm x \sim {\cal P}}\{ f(\bm x) \leq 0 \} \geq 1-\epsilon
	\end{equation}
	holds if and only if there exists $\bm Q \in \mathbb{H}^{N+1}$ and $\nu \in \mathbb{R}$ such that
	\begin{subequations}\label{eq:lemma_drb}
		\begin{align}
		0 \geq  \min_{\nu\in \mathbb{R},\bm Q\in \mathbb{H}^{N+1}}  & \nu + \epsilon^{-1} {\rm Tr}(\bm \Omega \bm Q), \label{eq:lemma_drb_a}\\
		{\rm s.t.}~	&  \bm Q \succeq \begin{bmatrix}
		\bm A & \bm b \\
		\bm b^H & c - \nu
		\end{bmatrix},\label{eq:lemma_drb_b} \\
		&~ \bm Q \succeq \bm 0, \label{eq:lemma_drb_c}
		\end{align}
	\end{subequations}
	where 
	\[  \bm \Omega  = \begin{bmatrix}
	\bm \Sigma + \bar{\bm \mu} \bar{\bm \mu}^H  & \bar{\bm \mu}\\
	\bar{\bm \mu}^H & 1
	\end{bmatrix}.   \]
\end{Theorem}  

By Theorem~\ref{lemma:drb}, the outage constraint~\eqref{eq:robust_sec_approx} is  satisfied if and only if 
\begin{subequations}\label{eq:drb_conic_eqv_1}
	\begin{align}
	0 \geq  \min_{\nu_1 \in \mathbb{R},\bm Q_1 \in \mathbb{H}^{N+1}} & \nu_1 + \epsilon_{\rm sec}^{-1} {\rm Tr}(\bm \Omega \bm Q_1), \label{eq:drb_conic_eqv_1a}\\
	{\rm s.t.} &~  \bm Q_1 \succeq {\cal G}(\bm W - \eta \bm \Phi, c_1-\nu_1),\label{eq:drb_conic_eqv_1b} \\
	&~ \bm Q_1 \succeq \bm 0, \label{eq:drb_conic_eqv_1c}
	\end{align}
\end{subequations}
where $c_1 = \sum_{k,\ell}[ {\cal M}(\bm W - \eta \bm \Phi)]_{k\ell} - { \sigma^2\eta }/{|\beta_1|^2}$ and
\[  {\cal G}(\bm X, x) \triangleq \begin{bmatrix}
{\cal L}( {\cal M}^{\Re}(\bm X) ) &  {\cal F}(  {\cal M}^{\Im}(\bm X)  )/2\\
{\cal F}(  {\cal M}^{\Im}(\bm X)  )^T/2 &  x
\end{bmatrix}.
\]

%\[{\cal G}(\bm W - \eta \bm \Phi, c_1-\nu_1) = \begin{bmatrix}
%{\cal L}( {\cal M}^{\Re}(\bm W - \eta \bm \Phi) ) &  {\cal F}(  {\cal M}^{\Im}(\bm W - \eta \bm \Phi)  )\\
%{\cal F}(  {\cal M}^{\Im}(\bm W - \eta \bm \Phi)  )^T &  c_1 - \nu_1
%\end{bmatrix} \] and $c_1 = \sum_{k,\ell}[ {\cal M}(\bm W - \eta \bm \Phi)]_{k\ell} - \frac{ \sigma^2 }{|\beta_1|^2}\eta$. 

Similarly,  the outage constraint~\eqref{eq:robust_est_eqv} is satisfied if
\begin{subequations}\label{eq:drb_conic_eqv_2}
	\begin{align}
	0 \geq  \min_{\nu_2 \in \mathbb{R},\bm Q_2 \in \mathbb{H}^{N+1}} & \nu_2 + \epsilon_{\rm est}^{-1} {\rm Tr}(\bm \Omega \bm Q_2), \label{eq:drb_conic_eqv_2a}\\
	{\rm s.t.} &~  \bm Q_2 \succeq - {\cal G}(\bm W+\bm \Phi, c_2+ \nu_2) ,\label{eq:drb_conic_eqv_2b} \\
	&~ \bm Q_2 \succeq \bm 0, \label{eq:drb_conic_eqv_2c}
	\end{align}
\end{subequations}
where $c_2 = \sum_{k,\ell}[ {\cal M}(\bm W+\bm \Phi)]_{k\ell} -\bar{\zeta}$.

%\[  {\cal G}_2(\bm W+\bm \Phi)= \begin{bmatrix}
%-{\cal L}( {\cal M}^{\Re}(\bm W+\bm \Phi) ) &  -{\cal F}(  {\cal M}^{\Im}(\bm W+\bm \Phi)  )\\
%-{\cal F}(  {\cal M}^{\Im}(\bm W+\bm \Phi)  )^T &  c_2 - \nu_2
%\end{bmatrix} \] and $c_2 = -\sum_{k,\ell}[ {\cal M}(\bm W+\bm \Phi)]_{k\ell} +\bar{\zeta}$.

Putting all the above pieces together, we can recast the robust SRM problem~\eqref{eq:main_robust}  as
\begin{subequations} \label{eq:main_robust_conic}
	\begin{align}
	\max_{\bm W, \bm \Phi, \bm Q_1, \bm Q_2, \nu_1, \nu_2, \eta} & ~ \frac{\sigma^2 + |\beta_2|^2 \bm h(\theta_2)^H (\bm W   + \bm \Phi) \bm h(\theta_2) }{(1+\eta)(\sigma^2 +  |\beta_2|^2 \bm h(\theta_2)^H \bm \Phi \bm h(\theta_2) )} \label{eq:main_robust_conic_a} \\
	{\rm s.t.} & ~ \nu_1 + \epsilon_{\rm sec}^{-1} {\rm Tr}(\bm \Omega \bm Q_1) \leq 0, \label{eq:main_robust_conic_b} \\
	& ~ \nu_2 + \epsilon_{\rm est}^{-1} {\rm Tr}(\bm \Omega \bm Q_2) \leq 0, \label{eq:main_robust_conic_c} \\
	& ~ \eqref{eq:drb_conic_eqv_1b}, \eqref{eq:drb_conic_eqv_1c}, \eqref{eq:drb_conic_eqv_2b}, \eqref{eq:drb_conic_eqv_2c}, \label{eq:main_robust_conic_d} \\
	& ~ {\rm Tr}(\bm W + \bm \Phi) \leq P, \label{eq:main_robust_conic_e} \\ 
	& ~ \bm W \succeq \bm 0, \quad \bm \Phi \succeq \bm 0,  \label{eq:main_robust_conic_f} \\
	& ~ 0\leq \eta \leq N |\beta_1|^2 P/\sigma^2, \label{eq:main_robust_conic_g} \\
	& ~ {\rm rank}(\bm W) \leq 1,  \label{eq:main_robust_conic_h}
	\end{align}
\end{subequations}
where the upper bound in~\eqref{eq:main_robust_conic_g} is deduced from~\eqref{eq:SINR_b}  as follows:
\begin{align*}
{\sf SINR}_{\rm eve} & = \frac{p_1 |\beta_1|^2 | \bm h(\theta_{1})^H \bm w |^2  }{ \sigma^2 + p_2 |\beta_1|^2 | \bm h(\theta_{1})^H \bm \varphi |^2  } \\
& \leq \frac{P |\beta_1|^2 | \bm h(\theta_{1})^H \bm w|^2}{\sigma^2}\\
& \leq N |\beta_1|^2 P/\sigma^2 ,
\end{align*}
where the inequality is achieved when $p_1 = P$, $p_2=0$ and $\bm w = \bm h(\theta_{1})/\| \bm h(\theta_{1})\|$. By dropping the rank-one constraint on $\bm W$ and applying Charnes-Cooper transformation, problem~\eqref{eq:main_robust_conic} can be relaxed as
\begin{subequations} \label{eq:main_robust_CCT}
	\begin{align}
	&\hspace{-5pt}	\max_{ \substack{\tilde{\bm W}, \tilde{\bm \Phi}, \tilde{\bm Q}_1, \tilde{\bm Q}_2, \\\tilde{\nu}_1, \tilde{\nu}_2, \eta, \kappa}}  ~ \sigma^2 \kappa + |\beta_2|^2 \bm h(\theta_2)^H (\tilde{\bm W}   + \tilde{\bm \Phi}) \bm h(\theta_2)  \label{eq:main_robust_CCT_a} \\
	& \qquad ~ 	{\rm s.t.}  ~ \sigma^2 \kappa +  |\beta_2|^2 \bm h(\theta_2)^H \tilde{\bm \Phi} \bm h(\theta_2)  = \frac{1}{1+\eta}, \label{eq:main_robust_CCT_b} \\
	& \qquad \qquad \tilde{\nu}_1 + \epsilon_{\rm sec}^{-1} {\rm Tr}(\bm \Omega \tilde{\bm Q}_1) \leq 0,  \label{eq:main_robust_CCT_c} \\
	& \qquad \qquad  \tilde{\nu}_2 + \epsilon_{\rm est}^{-1} {\rm Tr}(\bm \Omega \tilde{\bm Q}_2) \leq 0,  \label{eq:main_robust_CCT_d}\\
	& \qquad \qquad  \tilde{\bm Q}_1 \succeq {\cal G}(\tilde{\bm W} - \eta \tilde{\bm \Phi}, \tilde{c}_1 - \tilde{\nu}_1) ,  \label{eq:main_robust_CCT_e} \\
	& \qquad \qquad  \tilde{\bm Q}_2 \succeq -{\cal G}(\tilde{\bm W}+ \tilde{\bm \Phi}, \tilde{c}_2 + \tilde{\nu}_2) ,  \label{eq:main_robust_CCT_f} \\
	& \qquad \qquad {\rm Tr}(\tilde{\bm W} + \tilde{\bm \Phi}) \leq \kappa P  ,  \label{eq:main_robust_CCT_g} \\ 
	& \qquad \qquad   \tilde{\bm W}\succeq \bm 0,  \tilde{\bm \Phi} \succeq \bm 0,   \tilde{\bm Q}_1 \succeq \bm 0,  \tilde{\bm Q}_2 \succeq \bm 0,    \label{eq:main_robust_CCT_h} \\
	& \qquad \qquad  0\leq \eta \leq N |\beta_1|^2 P/\sigma^2, ~  \kappa \geq 0, \label{eq:main_robust_CCT_i}
	\end{align}
\end{subequations} 
where $\tilde{c}_1 = \sum_{k,\ell}[ {\cal M}(\tilde{\bm W }- \eta \tilde{\bm \Phi})]_{k\ell} - { \sigma^2 \eta  \kappa}/{|\beta_1|^2}$ and $\tilde{c}_2 =  \sum_{k,\ell}[ {\cal M}(\tilde{\bm W}+\tilde{\bm \Phi})]_{k\ell} - \kappa\bar{\zeta} $.

Notice that problem~\eqref{eq:main_robust_CCT} is still nonconvex due to the coupled variable $\eta \tilde{\bm \Phi}$ and the equality constraint~\eqref{eq:main_robust_CCT_b}. Nevertheless, for given $\eta$ problem~\eqref{eq:main_robust_CCT} is an SDP w.r.t. the remaining variables. Furthermore, since $\eta$ is within an interval (cf.~\eqref{eq:main_robust_CCT_i}), one can performing bisection over $\eta$ to iteratively solve problem~\eqref{eq:main_robust_CCT}. After solving problem~\eqref{eq:main_robust_CCT}, we can recover $\bm W = \tilde{\bm W}/\kappa$ and ${\bm \Phi} = \tilde{\bm \Phi}/\kappa$. If ${\rm rank}(\bm W) = 1$, then the beamformer $\bm w$ can be obtained by eigendecomposition; otherwise, Gaussian randomization may be used to generate the beamformer. Interestingly, our numerical experience suggests that solving problem~\eqref{eq:main_robust_CCT} always leads to a rank-one $\tilde{\bm W}$; i.e., no relaxation gap between problems~\eqref{eq:main_robust_conic} and \eqref{eq:main_robust_CCT}. So far, we are not able to give a rigorous proof for this observation and we leave this as a future work.

%\newpage
\section{Simulation Results} \label{sec:num_results}

We use Monte-Carlo simulations to evaluate the performance of the proposed designs. In our simulations, the complex path-loss coefficient $\beta_1$ at the target is given by $|\beta_1|^2 = \frac{G_{\rm t} G_{\rm_1} \lambda^2}{(4 \pi)^2 d_{\rm 1}^2}$,
%\begin{equation} \label{eq:path_loss_coeffcient}
%|\beta_1|^2 = \frac{G_{\rm t} G_{\rm_1} \lambda^2}{(4 \pi)^2 d_{\rm 1}^2}
%\end{equation}
where $d_{\rm 1}$ is the distance from DFRC transmitter to the target, $G_{\rm t}$ and $G_{\rm 1}$ denote the transmit antenna gain at the DFRC transmitter and the receive antenna gain at the target, respectively. The CU's $\beta_2$ is defined similarly. The complex combined gain $\beta_3$ is given by $|\beta_3|^2 = \frac{G_{\rm t}^2 \lambda^2 S}{(4 \pi)^3 d_{\rm 1}^4}$,
%\begin{equation} \label{eq:combined_gain}
%|\beta_3|^2 = \frac{G_{\rm t}^2 \lambda^2 S}{(4 \pi)^3 d_{\rm 1}^4}
%\end{equation}
where $S$ is radar cross-section (RCS). The results to be presented in this section are based on the following settings: The number of  antennas at  the DFRC transmitter is $N=10$; the noise's variance $\sigma^2$ is $0$\,dB relative to $|\beta_1|^2$; the channel bandwidth $B=10$\,MHz; the pulse duration $T=10$\,$\mu$s; the duty factor $\delta=0.5$; the carrier wavelength $\lambda=0.1$\,m; the inter-antenna spacing $d={\lambda}/{4}$, the variance of target process $\sigma^2_{\tau,{\rm proc}}=4.44 \times 10^{-15}$, the scaling constant $\gamma^2={(2 \pi)^2}/{12}$; $G_{\rm t}=10$\,dB, $G_{\rm 1}=G_{\rm 2}=0$\,dB, $S=1$, $d_{\rm 1}=1$\,Km, $d_{\rm_2}=0.5$\,Km, the angle of departure from the DFRC transmitter to the target $\theta_{\rm 1}=30^\circ$ and $|\theta_2- \theta_{1}|= \Delta \theta$.

\subsection{The No-AN Case}
\begin{figure}[!htbp]
	\includegraphics[width=8cm]{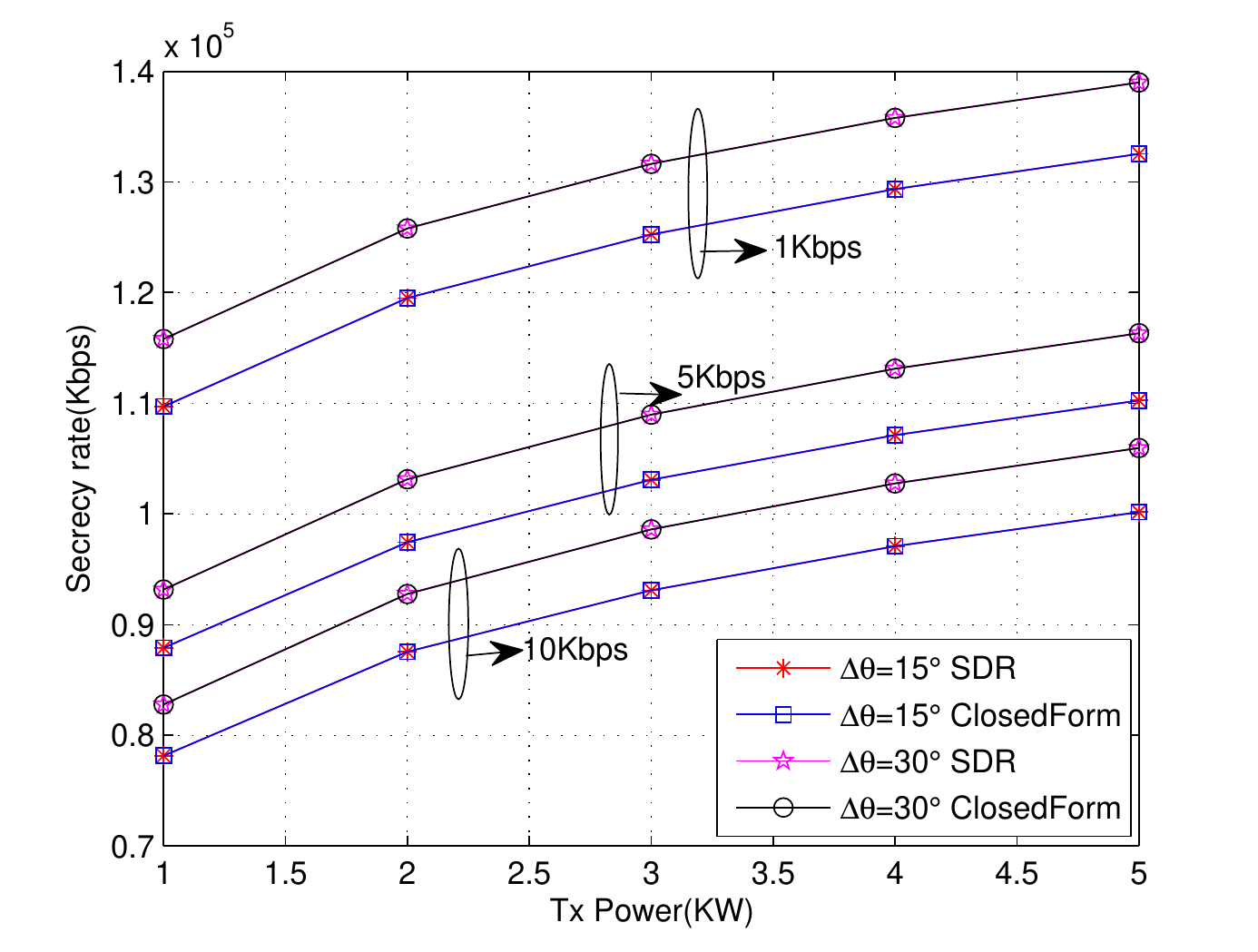}
	\caption{Secrecy rate vs. transmit power.}
	\label{SR_vs_power}
\end{figure}

\begin{figure}[!htbp]
	\includegraphics[width=8cm]{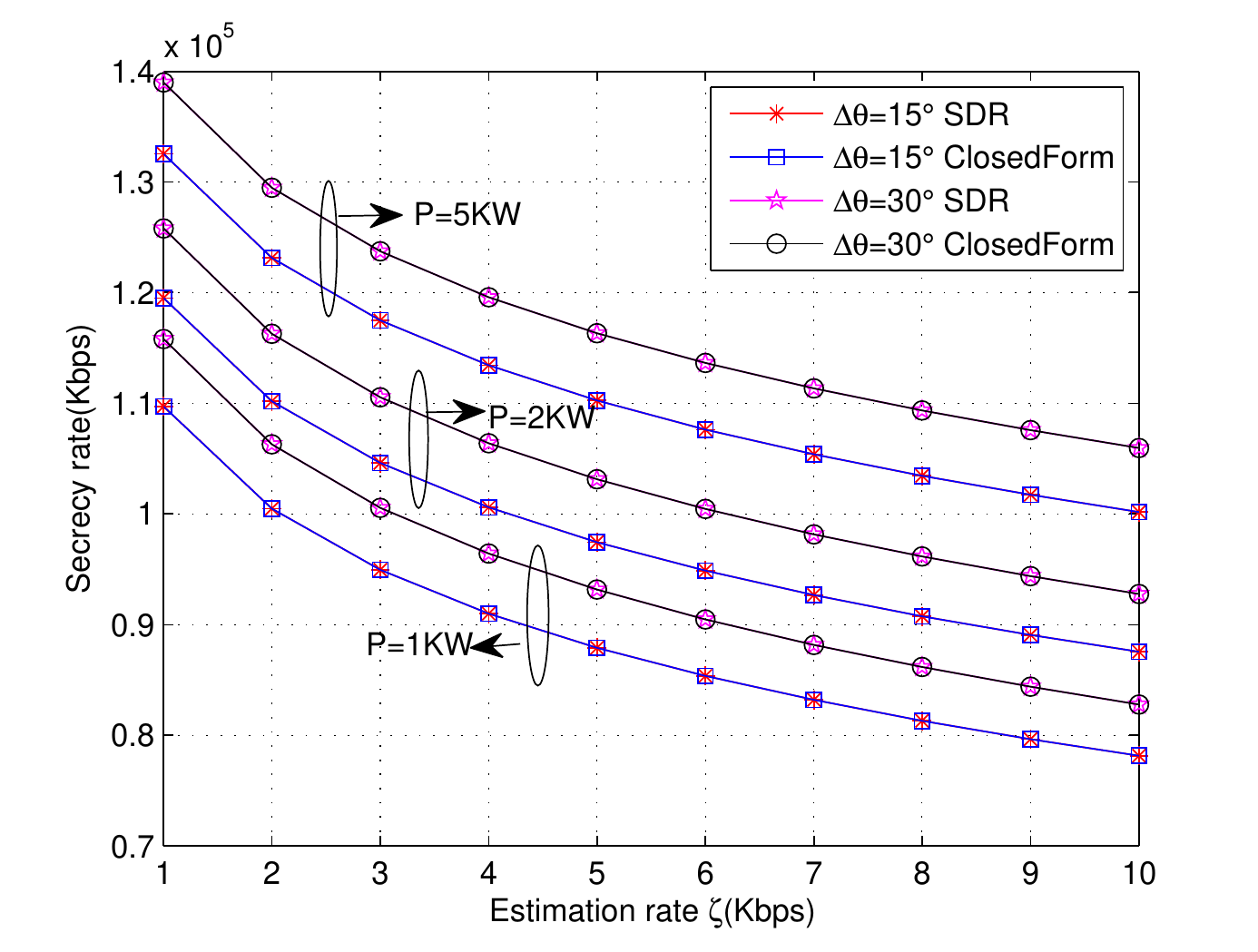}
	\caption{Secrecy rate vs. the estimation rate.}
	\label{SR_vs_est_rate}
\end{figure}

Figures~\ref{SR_vs_power} and \ref{SR_vs_est_rate} compare the secrecy rate performance of the closed-form solution~\eqref{eq:opt_beamer_expression} and the SDR for different transmit power and estimation rate, respectively. As seen, both schemes achieve the same secrecy rate, which corroborates the SDR tightness in Claim~\ref{fact}. In addition, larger transmit power or lower estimation rate requirement gives rise to higher secrecy rate, because of the enlarged feasible set of problem~\eqref{eq:main}. Moreover, with the increase of $\Delta \theta$, the secrecy performance is improved. This is  because larger $\Delta \theta$ reduces the  correlation between the CU's channel and the eavesdropping channel, or in other words makes the two channels more distinguishable. Consequently, it favors the DFRC transmitter to generate a discriminative beampattern towards the direction of the CU and the target, so that the reception of the CU is improved whereas that of the target is crippled.

\begin{figure}[!htbp]
	\includegraphics[width=8cm]{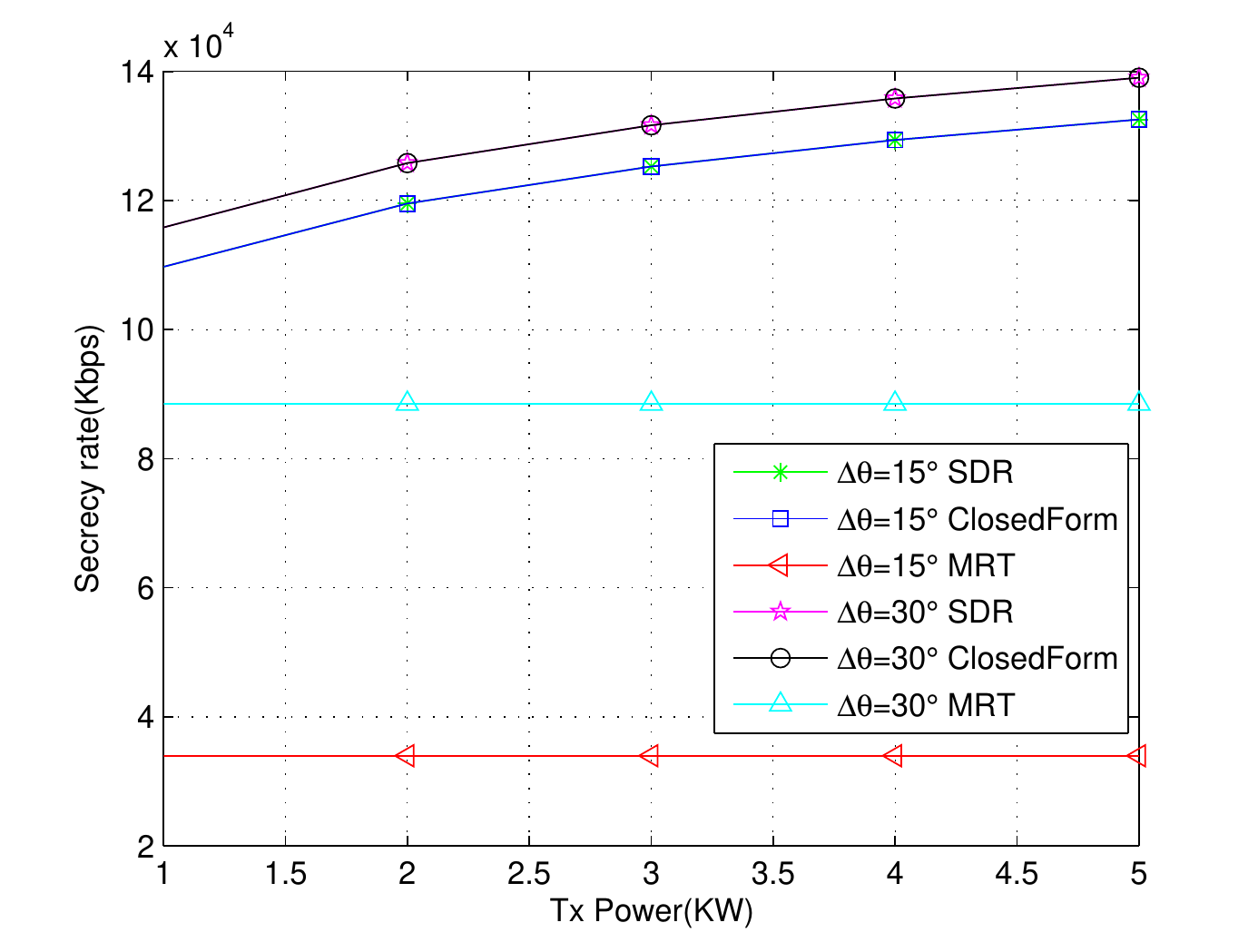}
	\caption{Comparison of closed-form solution and MRT under different transmit power $P$ ($\zeta
		=1\,$Kbps).}
	\label{fig:sdr_vs_mrt_1}
\end{figure}

\begin{figure}[!htbp]
	\includegraphics[width=8cm]{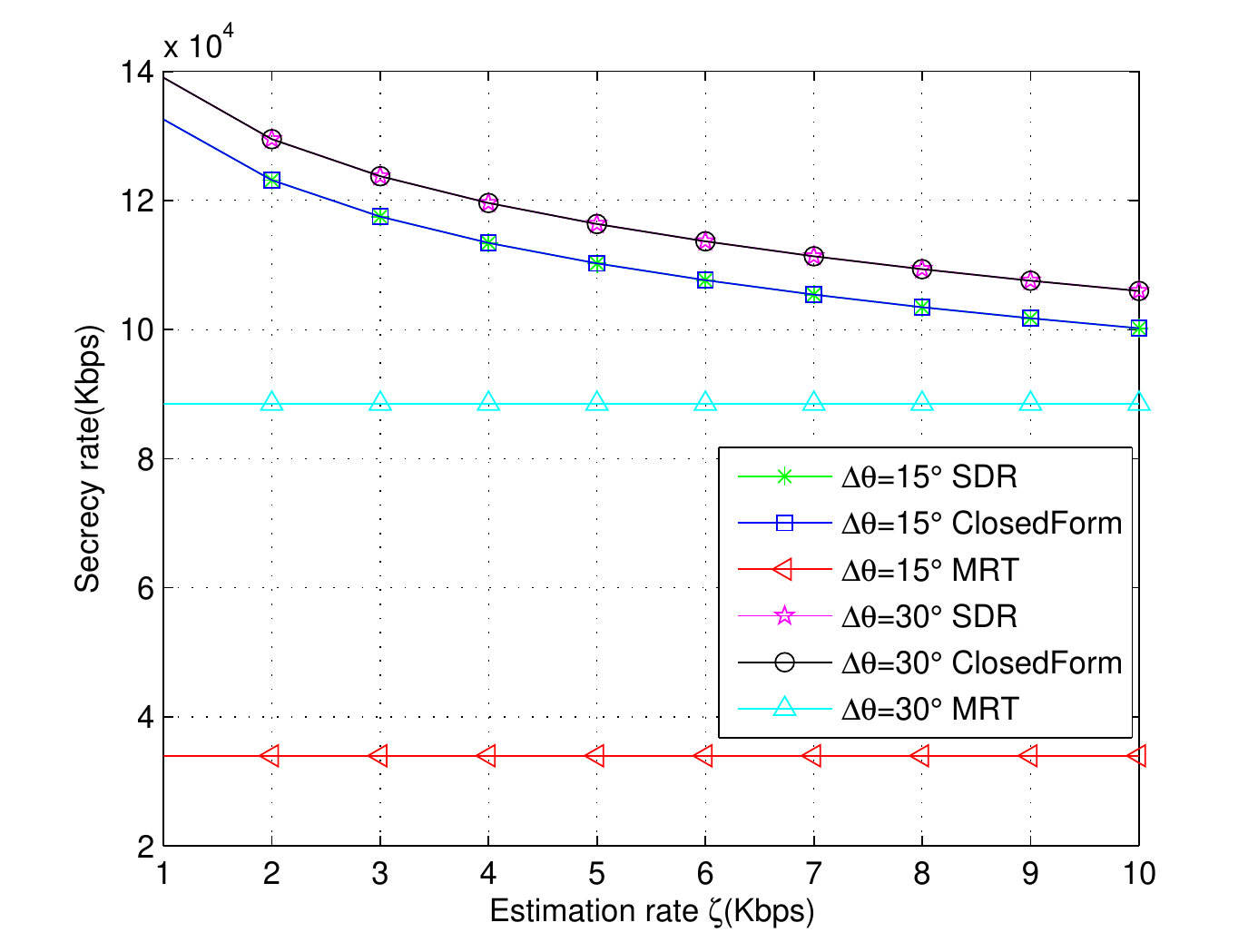}
	\caption{Performance comparison of closed-form solution and MRT under different estimation rate ($P=5$\,KW).}
	\label{fig:sdr_vs_mrt_2}
\end{figure}

To demonstrate the superior performance of the proposed design, we compare the proposed closed-form solution and a simple benchmark transmit scheme---the maximum ratio transmission (MRT),  which beamforms towards the direction of CU with full power.  Figures~\ref{fig:sdr_vs_mrt_1} and \ref{fig:sdr_vs_mrt_2} show the results.  It is seen from Figure~\ref{fig:sdr_vs_mrt_1} that with the increase of the transmit power, the secrecy rate of MRT is almost flattened. This is because as the power increases, MRT improves CU's reception and meanwhile also improves the target's reception, thereby leading to marginal increase of the secrecy rate, especially for the considered high-power transmit region. Similarly, Figure~\ref{fig:sdr_vs_mrt_2} shows that the MRT is almost invariant with  the change of $\zeta$. This is because for the considered range of $\zeta$, the estimation rate is already over satisfied and consequently, the secrecy rate is invariant when MRT is used with full transmit power.

\subsection{The AN-aided Case}

\begin{figure}[!htbp]
	\includegraphics[width=8cm]{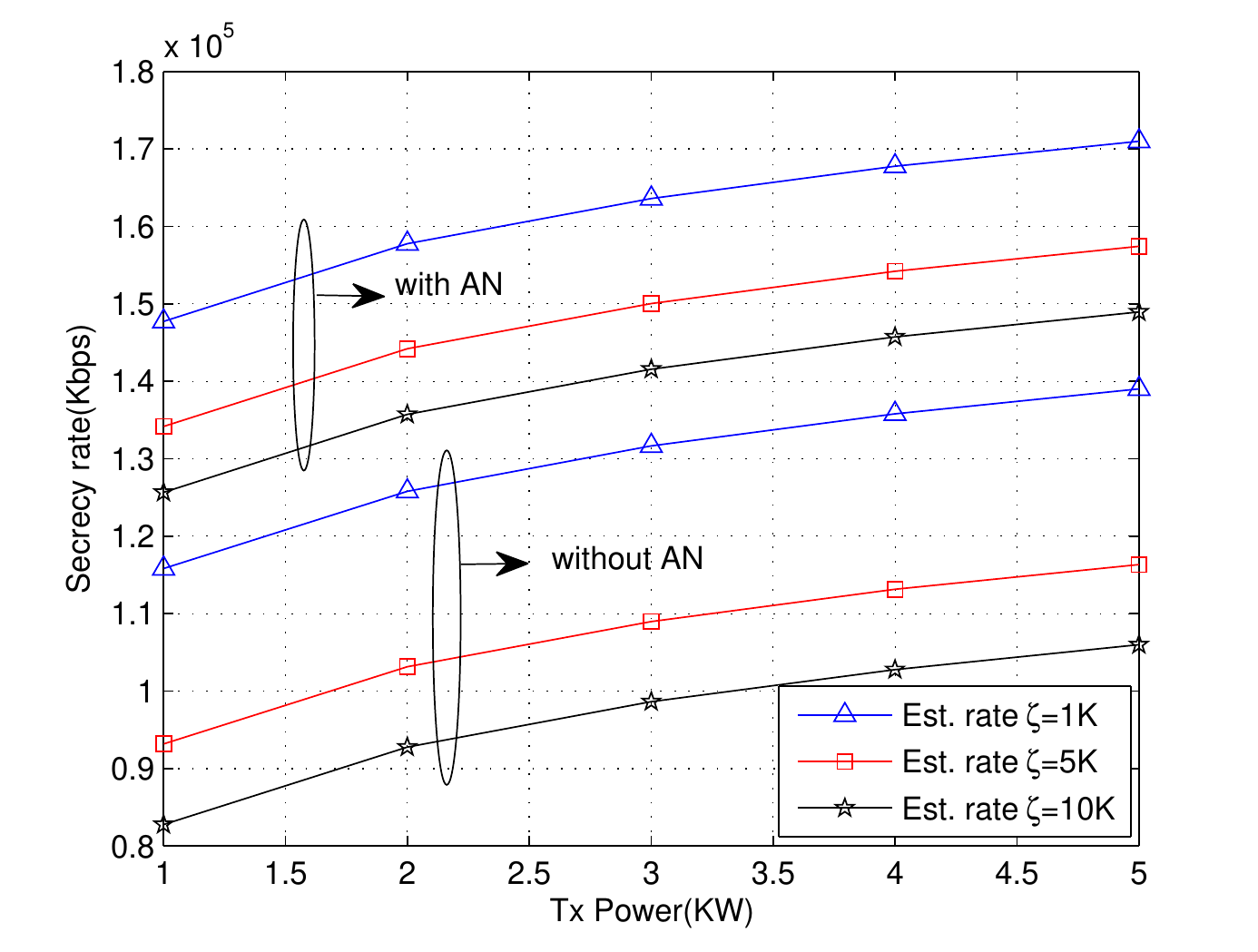}
	\caption{Comparison of  no-AN  and AN-aided transmit schemes. ($\theta_1=30^\circ$, $\theta_2=60^\circ$).}
	\label{SR_vs_Pwr_AN}
\end{figure}

Let us first compare the secrecy performance with and without AN. Figure~\ref{SR_vs_Pwr_AN} shows the secrecy rate against the transmit power for different estimation rate threshold. From the figure, we see that AN-aided transmit scheme outperforms the no-AN scheme. Especially, when the estimation rate requirement becomes more stringent, say from 1Kbps to 10Kbps, the performance gain of using AN is more prominent. This demonstrates the dual benefits of AN in securing information and helping radar estimation.  More specifically, as we mentioned in Introduction, while the AN behaves like random noise for the target, it is  deterministically known for the DFRC transmitter as a prior. Hence, the echo of the AN can be further exploited for the radar purpose. Meanwhile, since AN is treated as interference at the target, it also help prevent interception, and thus attains higher secrecy rate than the no-AN case.

\begin{figure}[!h]
	\includegraphics[width=8cm]{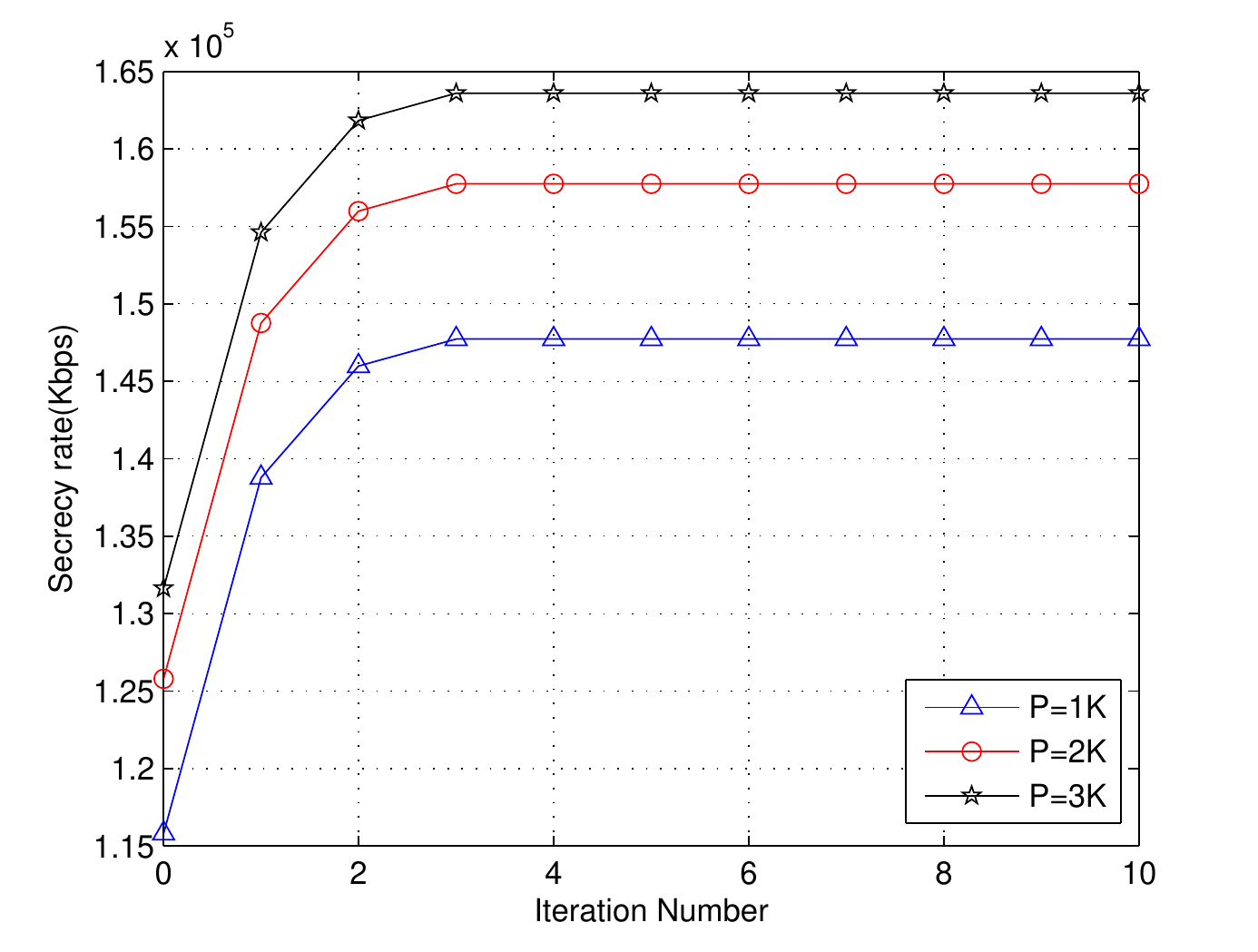}
	\caption{Convergence behavior of the AN-aided design ($\zeta=1$Kbps, $\theta_1=30^\circ$, $\theta_2=60^\circ$).}
	\label{SR_vs_IterNum_AN}
\end{figure}

Figure~\ref{SR_vs_IterNum_AN} shows the convergence behavior of the AO Algorithm~\ref{AL1}. It is seen that the AO algorithm converges fast usually within three iterations for the considered scenarios. 

\begin{figure}[!htbp]
	\includegraphics[width=8cm]{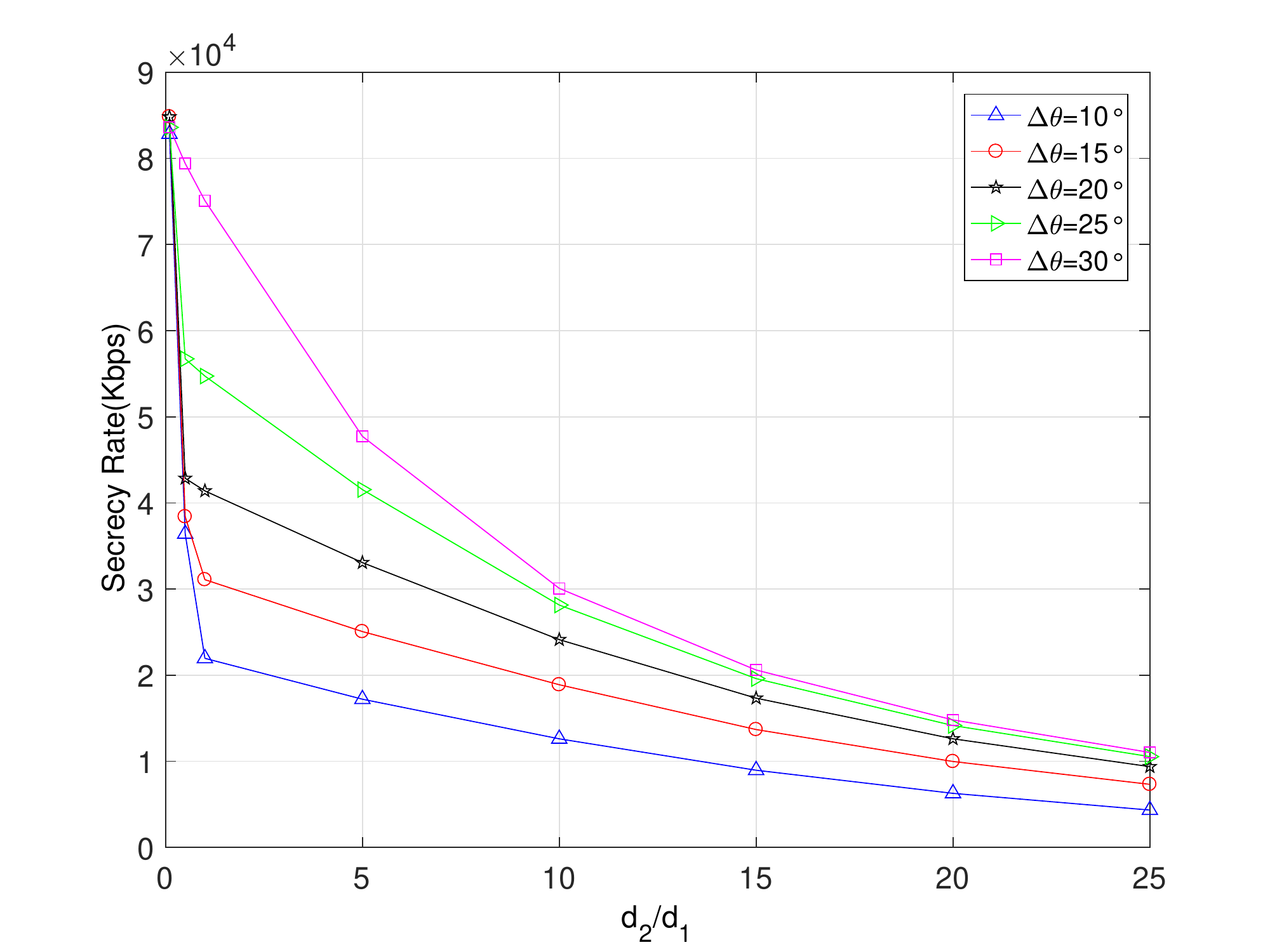}
	\caption{Secrecy rate vs.  distance ratio $d_2/d_1$ ($d_1=1\,$Km, $P=100$W, $\zeta=20$Kbps, $\theta_1=30^\circ$).}
	\label{SR_vs_Dist}
\end{figure}

Next, we investigate how the distance ratio $d_2/d_1$ and the angle difference $\Delta \theta$ affect the secrecy rate and the power allocation of the AN. The ratio $d_2/d_1$  and the $\Delta \theta$  reflect the relative channel strength and the channel correlation between the CU and the target, respectively. Figure~\ref{SR_vs_Dist} plots the secrecy rate vs. the ratio $d_2/d_1$ with $d_1$ fixed at $1$\,Km. As the CU moves away from the DFRC transmitter, the secrecy rate is decreased because the path-loss becomes larger and the CU's channel capacity is decreased. Also, for the fixed ratio $d_2/d_1$, the secrecy rate is improved when the $\Delta \theta$ is large. This is consistent with the no-AN case  (cf.~Figures~\ref{SR_vs_power} and \ref{SR_vs_est_rate}), since larger $\Delta \theta$ leads to less correlated channels, which makes CU and the target more discriminative in the spatial domain, that is, the DFRC transmitter can more easily generate discriminative beampattern to improve the CU's reception and suppress the target's interception.

\begin{figure}[!htbp]
	\includegraphics[width=8cm]{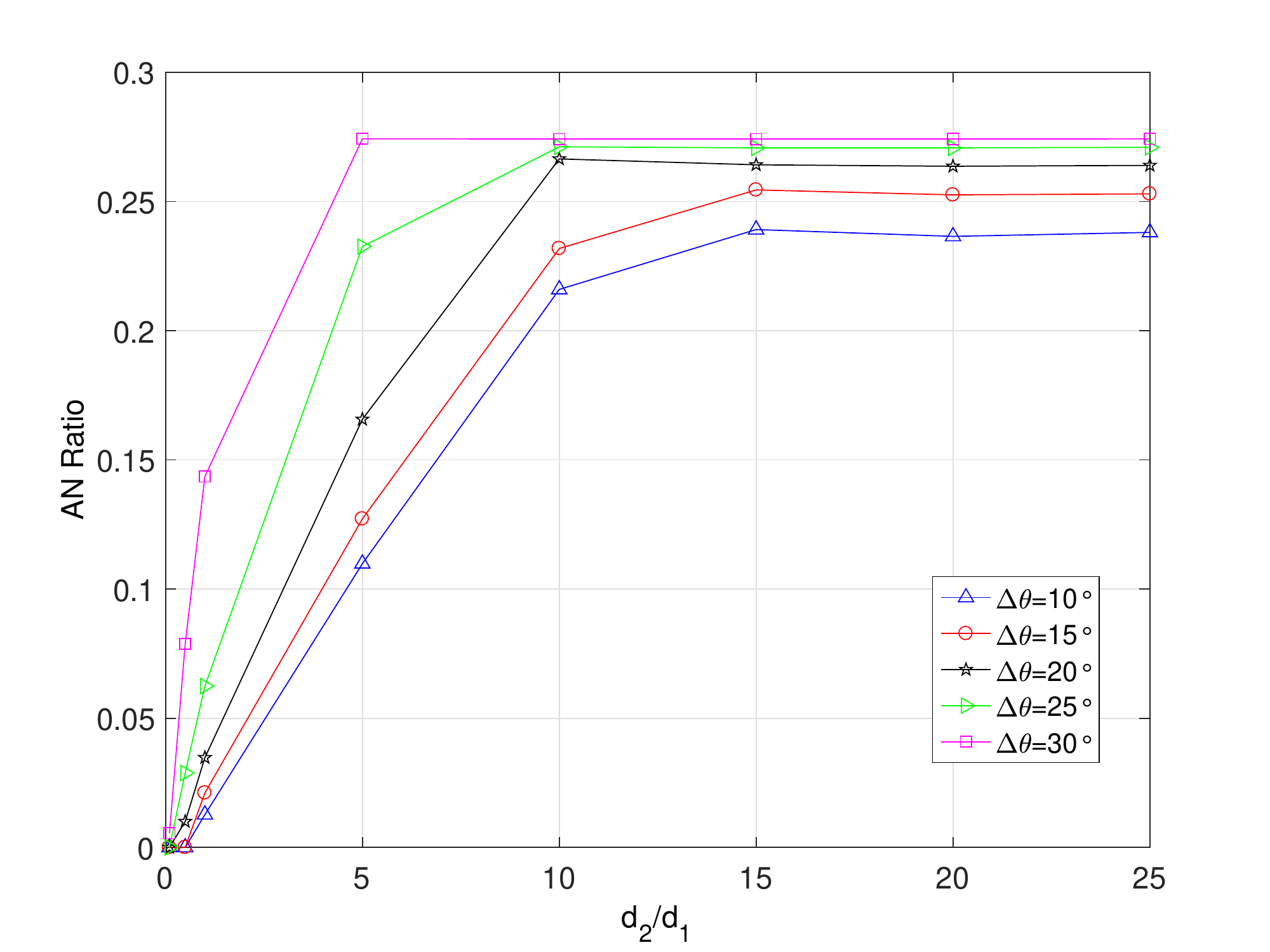}
	\caption{AN power ratio vs. distance ratio $d_2/d_1$  ($d_1=1\,$Km, $P=100$W, $\zeta=20$Kbps, $\theta_1=30^\circ$).}
	\label{Ratio_vs_Dist}
\end{figure}

Figure~\ref{Ratio_vs_Dist} plots the ratio of the AN's power to the total transmit power against the distance ratio $d_2/d_1$ under the same setting as Figure~\ref{SR_vs_Dist}. From the figure, we have the following observations: Firstly, as the ratio $d_2/d_1$ becomes larger or the CU's channel becomes weaker, we need to allocate more power to AN. This can be explained as follows. Typically, there are two ways to improve the secrecy rate: one is to degrade the eavesdropper's reception, say by jamming, and the other is to improve the CU's reception with large transmit power. However, when the CU's channel is weak, the former is more power efficient than the latter, and vice versa. Therefore, for large $d_2/d_1$, it prefers to exploit  AN  to improve the secrecy rate. Secondly, when $\Delta \theta$ increases, more power should be allocated to the AN. This is because when the spatial channels of the CU and the target become more discriminative, it is hard to leverage only on a single beamformer to cater for both the estimation task and the secret communication task. Therefore, by allocating more power on the AN, the estimation and the communication can be addressed with AN and beamforming, respectively.

\subsection{The Imperfect CSI Case}
\begin{figure}[!htbp]
	%	\centering
	\includegraphics[width=8cm,height=6.5cm]{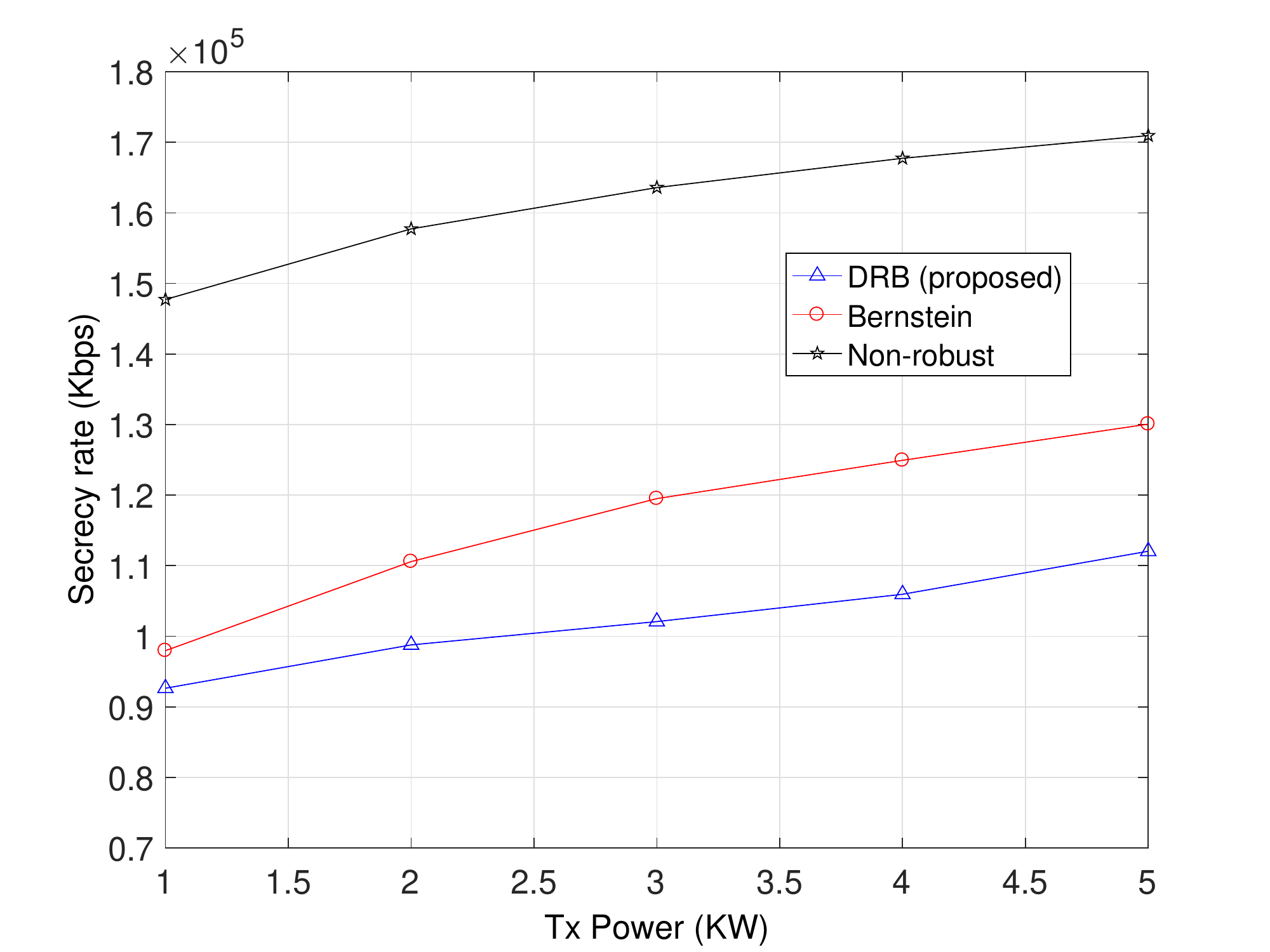}
	%	\vspace{-10pt}
	\caption{Robust secrecy rate vs. transmit power $P$ ($\zeta=1$Kbps, $d_{\rm 1}=1$\,Km, $d_{\rm 2}=0.5$\,Km, $\theta_1=30^\circ$, $\theta_2=60^\circ$).}
	\label{SR_vs_Pwr_robust}
	%	\vspace{-20pt}
\end{figure}

In this subsection, we consider the imperfect CSI case. The simulation settings are basically identical to the perfect CSI case. The outage probabilities of secrecy rate and estimation rate threshold are set to $\epsilon_{\rm sec}=\epsilon_{\rm est}=1\%$. As for the channel phase error model, we set the mean $\bar{\bm \mu}=\bm 0_{N}$, the variance $\bm \Sigma=(5 \pi/180)^2  \bm I_{N}$.
Figure~\ref{SR_vs_Pwr_robust} plots the secrecy rate against the transmit power $P$ under various designs. In the legend, ``Non-robust'' design represents the AN-aided design  based on the estimated $\bar{\theta}_{1}$ without considering the phase error;
``Bernstein'' corresponds to the robust design in~\cite{So16}, which assumes Gaussian distribution of the phase error and makes use of the Bernstein-type inequality to get a safe approximation of the outage probability constraints. We should mention that in~\cite{So16}, the  Bernstein-type inequality-based robust design is developed for multiuser multicast  beamforming, not for the DFRC application. Nevertheless, one can easily adapt the method in~\cite{So16} to the DFRC case. ``DRB (proposed)'' represents the proposed distributionally robust beamforming design in this paper. From the figure, we see that ``Non-robust'' attains the best rate performance, then followed by ``Bernstein'' and the ``DRB''. This is natural, since the first two designs either does not consider phase error or assumes the distribution set $\mathscr{D}(\bar{\bm \mu}, \bm \Sigma)$ is a singleton of Gaussian distribution. In either case, the resultant robust SRM problem  can be  regarded as a  relaxation of problem~\eqref{eq:main_robust}, thereby attaining higher secrecy rate than ``DRB''.

\begin{figure}[!htbp]
	%	\centering
	\includegraphics[width=8cm,height=6.5cm]{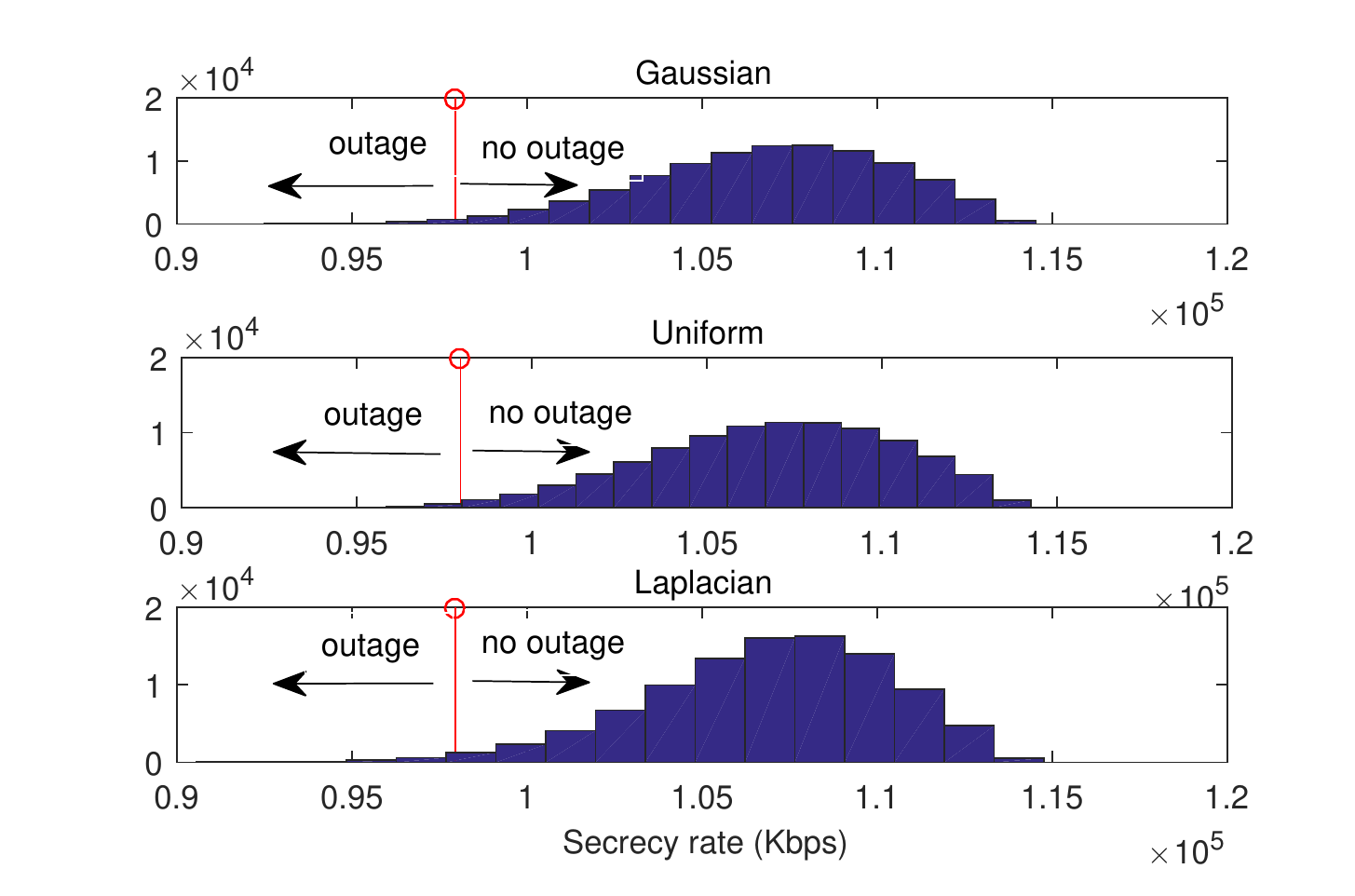}
	%	\vspace{-20pt}
	\caption{Empirical secrecy rate histograms of ``Bernstein'' under different error distributions ($P=1$\,KW, $\zeta=1$Kbps, $d_{\rm 1}=1$\,Km, $d_{\rm 2}=0.5$\,Km, $\theta_1=30^\circ$, $\theta_2=60^\circ$).}
	\label{Berstein_hist}
\end{figure}

\begin{figure}[!h]
	%	\centering
	\includegraphics[width=8cm,height=6.5cm]{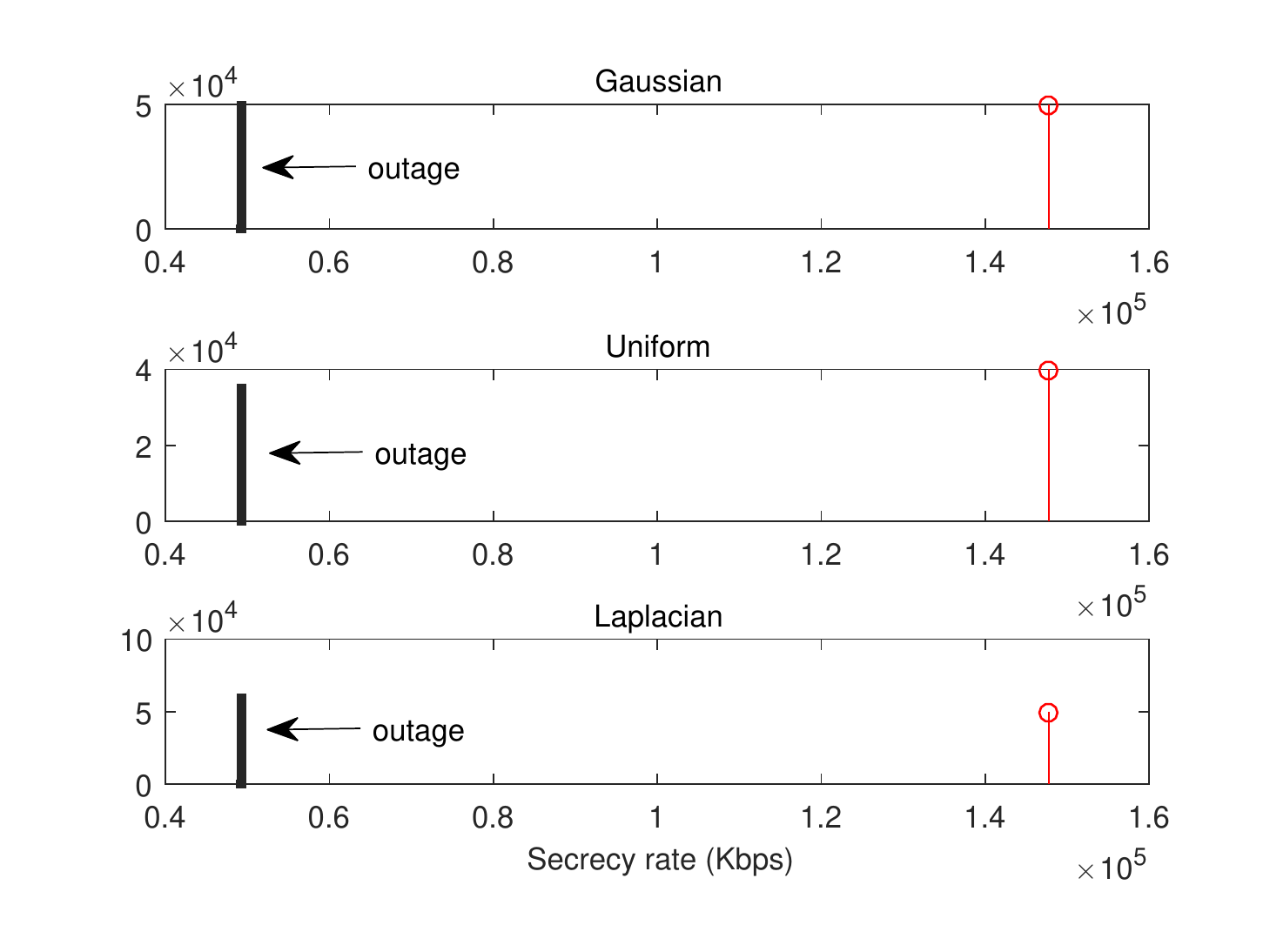}
	%	\vspace{-20pt}
	\caption{Empirical secrecy rate histograms of nonrobust design  under different error distributions ($P=1$\,KW, $\zeta=1$Kbps, $d_{\rm 1}=1$\,Km, $d_{\rm 2}=0.5$\,Km, $\theta_1=30^\circ$, $\theta_2=60^\circ$).}
	\label{nonrobust_hist}
	%	\vspace{-20pt}
\end{figure}

\begin{figure}[!h]
	%	\centering
	\includegraphics[width=8cm,height=6.5cm]{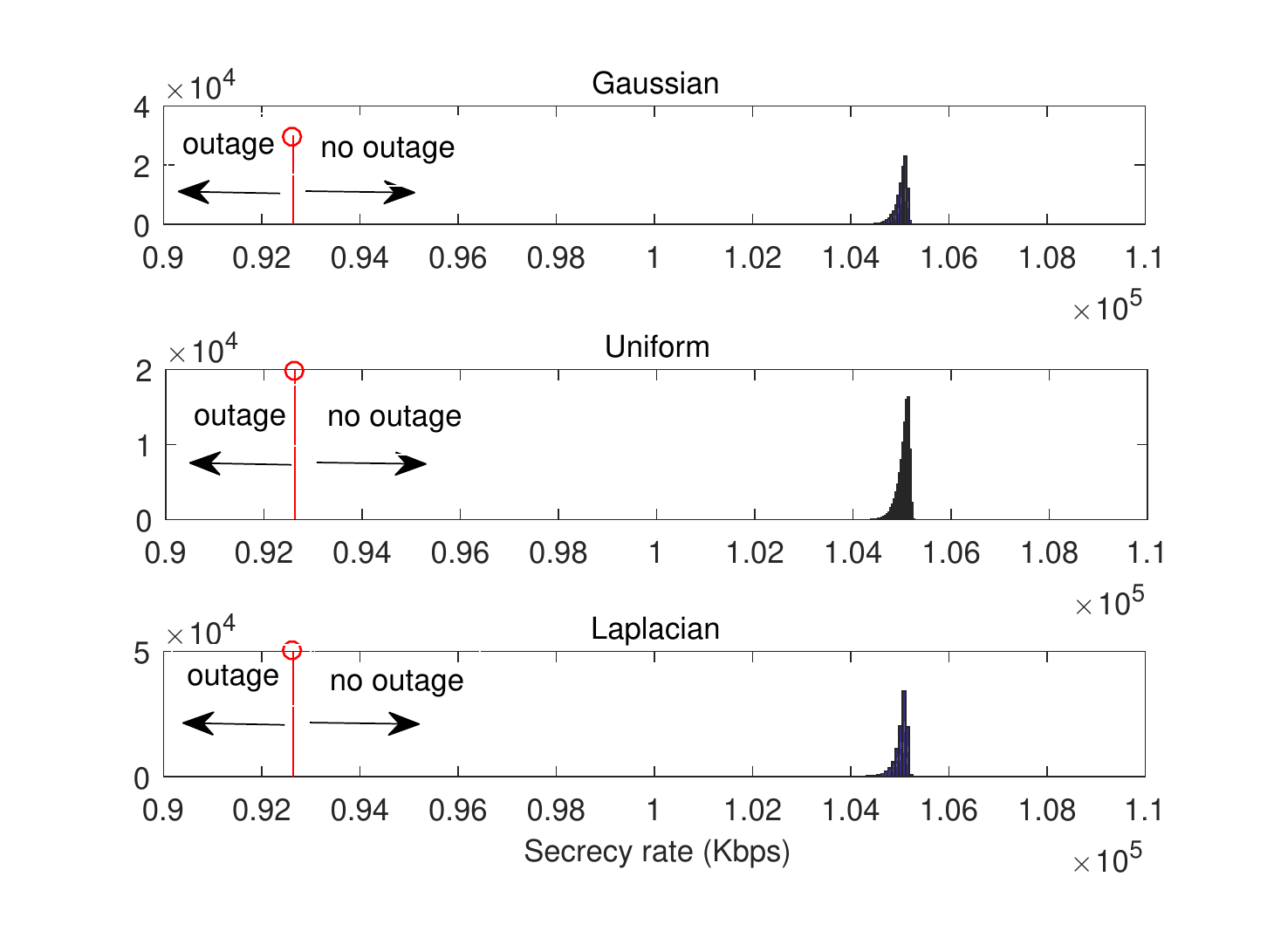}
	%	\vspace{-20pt}
	\caption{Empirical secrecy rate histograms of DRB under different error distributions ($P=1$\,KW, $\zeta=1$Kbps, $d_{\rm 1}=1$\,Km, $d_{\rm 2}=0.5$\,Km, $\theta_1=30^\circ$, $\theta_2=60^\circ$).}
	\label{DRB_hist}
	%	\vspace{-20pt}
\end{figure}

However, under the considered moment-based error model, the distribution of the phase error is not limited to Gaussian. Therefore, it is expected that the non-robust design and the Bernstein robust design based on Gaussian error assumption may not satisfy the outage probability  constraints~\eqref{eq:main_robust_b} and \eqref{eq:main_robust_c}. To verify this, we empirically evaluate the secrecy outage probability of  the three designs by generating random  realizations of different error distributions. Since it is unlikely to generate all the distributions in $\mathscr{D}(\bar{\bm \mu}, \bm \Sigma)$, we consider three representative
distributions, namely, Gaussian distribution, Uniform distribution and Laplacian distribution. We randomly generate $10^5$ phase error realizations for each of the three distributions, and calculate the secrecy rate $R_{\rm sec}^{j,t}$ for the $t$th realization of the type $j = 1,2,3$ distribution. After obtaining all $R_{\rm sec}^{j,t}$ for each $j$, we plot the histograms of $R_{\rm sec}^{j,t}$ for $j = 1,2,3$. Figure~\ref{Berstein_hist} shows the histograms of $R_{\rm sec}^{j,t}$ for the Bernstein design, where the red circled line corresponds to the theoretically computed secrecy rate from solving the robust SRM with the Bernstein method. If the empirical secrecy rate falls below the value at the red circled line, it means outage occurs. From the figure, we see that for more than $1\%$ error realizations, the empirical secrecy rate is below red circled line; that is, the ``Bernstein'' design cannot guarantee the required outage probability requirement, and transmission security is compromised.  Figure~\ref{nonrobust_hist} shows result for the non-robust design. It is seen that secrecy outage happens for all the error realizations, due to ignoring the phase error. By contrast,  Figure~\ref{DRB_hist} shows that all the empirical secrecy rates of the ``DRB'' are above the theoretically computed secrecy rate threshold, and thus no secrecy outage happens under the tested error distributions. 

\section{Conclusions} \label{sec:conclusions}
This paper has considered the joint communication and radar beamforming for DFRC system, when the target is potentially an eavesdropper. Three secrecy rate maximization (SRM) problems are formulated under both perfect and imperfect CSI of the target. For the perfect CSI case, we develop a closed-form solution to the SRM problem when only information beamforming is exploited. Upon the closed-form solution, an alternating optimization (AO) algorithm is proposed for the more challenging artificial-noise-aided SRM problem. For the imperfect CSI case, we considered a moment-based random phase error model and formulated a distributionally robust SRM problem. By leveraging on the semidefinite relaxation (SDR) and conic representation of the robust chance constraints, a tractable solution is obtained. Simulation results verify the efficacy and robustness of the proposed designs. Finally, we should mention that the focus of this work is on the secrecy-estimation rate and aiming at developing some closed-form and interpretable solutions to the DFRC beamforming by considering a single communication user and target. Nevertheless, the SRM formulation and the approach  developed in this work can be extended to other cases. For example, one may consider maximizing the estimation rate at the target with pre-specified  secrecy rate, or maximizing weighted sum of the estimation rate and the secrecy rate, or solving the SRM problem under more complex scenarios, e.g., multiple communication users and targets. For the latter case, the SDR tightness and the closed-form solution may not exist and we leave this as a future work.

\section{Appendix}
\subsection{Proof of Claim~\ref{ob:1}} \label{app:A}
Any optimal $\bm w^\star$ can be represented as
\[ \bm w^\star = \lambda_1^\star \bm u_1 + \lambda_2^\star \bm u_2 +  \bm U^{\perp} \bm x^\star \]
for some $\lambda_1^\star\in \mathbb{C}$, $\lambda_2^\star\in \mathbb{C}$ and $\bm x^\star \in \mathbb{C}^{N-2}$, where $\bm U^{\perp} \in \mathbb{C}^{N\times (N-2)}$ is the orthogonal complement of $[\bm u_1 ~\bm u_2]$. Let us show $\bm x^\star =\bm 0$ must hold at the optimal $\bm w^\star$ by contradiction. Suppose $\bm x^\star \neq \bm 0$. Then, $0< \rho = \| \bm U^{\perp} \bm x^\star \|^2 <1$ due to $\| \bm w^\star\|^2 = 1$. Construct $\bm w_0 =\lambda_1^\star \bm u_1 + \lambda_2^\star \bm u_2 $ and $\tilde{\bm w} = \frac{1}{\sqrt{1-\rho}} \bm w_0$. It can be verified that $\|\tilde{\bm w} \|^2=1$. In addition,  $\tilde{\bm w}^H \bm h_1 \bm h_1^H \tilde{\bm w} = \frac{1}{1-\rho} \|\bm w_0^H \bm h_1\|^2 = \frac{1}{{1-\rho}} \|(\bm w^\star)^H \bm h_1 \|^2 \geq \frac{\alpha}{{1-\rho}} > \alpha$, where the second equality is due to $\bm U^{\perp} \perp \bm h_1$. Hence, $\tilde{\bm w}$ is a feasible solution. Next, we show that $\tilde{\bm w}$ can attain higher objective value than $\bm w^\star$, thereby contradicting the optimality of $\bm w^\star$. Denote the objective of problem~\eqref{eq:main_fqcqp} as $f(\bm w)$. It holds that 
\[ f(\bm w^\star) = f(\bm w_0)< f(\frac{1}{\sqrt{1-\rho}} \bm w_0) = f(\tilde{\bm w})  \] 
where the first equality is because  $\bm U^{\perp}$ is the orthogonal complement of $[\bm u_1 ~ \bm u_2]$ (or $[\bm h_1 ~ \bm h_2]$), and the inequality is because $g(x ) \triangleq f(x \bm w_0)$ is strictly increasing w.r.t. $x \geq 1$, when $f(\bm w_0)>1$, i.e., positive secrecy rate is attained at $\bm w_0$ (or $\bm w^\star$). However, the above inequality contradicts with the optimality of $\bm w^\star$.

\subsection{Proof of Claim~\ref{ob:2}} \label{app:ob2}
The proof consists of three steps.

{\it Step 1}: Suppose that $(\hat{p}_1, \hat{p}_2, \hat{\bm w}, \hat{\bm \Phi})$ is an optimal solution of problem~\eqref{eq:main_AN}. Let 
$\hat{\mu}_1 =\bm h(\theta_{1})^H \hat{\bm \Phi} \bm h(\theta_{1}) $ and $\hat{\mu}_2 =\bm h(\theta_{2})^H \hat{\bm \Phi} \bm h(\theta_{2}) $. Now, consider the following problem:
\begin{subequations}\label{eq:proof_ob2_main}
	\begin{align}
	\min_{\bm \Phi\succeq \bm 0 } & ~ {\rm Tr}(\bm \Phi)  \label{eq:proof_ob2_main_a} \\
	{\rm s.t.} & ~ \bm h(\theta_{1})^H {\bm \Phi} \bm h(\theta_{1}) \geq \hat{\mu}_1, \label{eq:proof_ob2_main_b} \\
	& ~ \bm h(\theta_{2})^H {\bm \Phi} \bm h(\theta_{2}) \leq \hat{\mu}_2. \label{eq:proof_ob2_main_c} %\\
	%& ~ \bm \Phi \succeq \bm 0 \label{eq:proof_ob2_main_d}
	\end{align}
\end{subequations}
We show that any optimal solution $\bm \Phi^\star$ of problem~\eqref{eq:proof_ob2_main}, together with $(\hat{p}_1, \hat{p}_2, \hat{\bm w})$, is also optimal for problem~\eqref{eq:main_AN}. It follows from~\eqref{eq:proof_ob2_main_b} and \eqref{eq:proof_ob2_main_c} that 
\begin{equation} \label{eq:proof_ob2_2}
g(\hat{p}_1, \hat{p}_2, \hat{\bm w}, \bm \Phi^\star)\geq g(\hat{p}_1, \hat{p}_2, \hat{\bm w}, \hat{\bm \Phi})
\end{equation}
holds, and that the estimation rate constraint~\eqref{eq:main_AN_b} is satisfied. The remaining is to show that ${\rm Tr}(\bm \Phi^\star)=1$. Clearly, $\hat{\bm \Phi}$ is a feasible solution of problem~\eqref{eq:proof_ob2_main}. Hence, 
\[ {\rm Tr}(\bm \Phi^\star) \leq {\rm Tr}(\hat{\bm \Phi}) = 1. \]
Suppose  ${\rm Tr}(\bm \Phi^\star) <1$. Then, we can construct another $\tilde{\bm \Phi} = \bm \Phi^\star + \gamma \bm x \bm x^H $ for some $\gamma>0$ and $\bm x \in \mathbb{C}^N$ such that 
\begin{equation}\label{eq:proof_ob2_3}
\bm x^H \bm h(\theta_2) = 0, ~\bm x^H \bm h(\theta_1) \neq  0, ~ {\rm Tr}(\tilde{\bm \Phi})=1,
\end{equation}
where we have used the assumption  $\theta_{1} \neq \theta_{2}$. It is easy to see that $\tilde{\bm \Phi} $ is feasible for problem~\eqref{eq:proof_ob2_main} and that $g(\hat{p}_1, \hat{p}_2, \hat{\bm w}, \tilde{\bm \Phi}) > g(\hat{p}_1, \hat{p}_2, \hat{\bm w}, \bm \Phi^\star)$, which together with~\eqref{eq:proof_ob2_2} and \eqref{eq:proof_ob2_3} implies that $(\hat{p}_1, \hat{p}_2, \hat{\bm w}, \tilde{\bm \Phi})$ can attain higher objective value for problem~\eqref{eq:main_AN} than $(\hat{p}_1, \hat{p}_2, \hat{\bm w}, \hat{\bm \Phi})$; this contradicts with the optimality of $(\hat{p}_1, \hat{p}_2, \hat{\bm w}, \hat{\bm \Phi})$. Therefore, we must have ${\rm Tr}(\bm \Phi^\star) =1$. 

{\it Step 2:} Next, we show that any optimal solution of problem~\eqref{eq:proof_ob2_main} must be rank-one. Let $\zeta_1 \geq 0$, $\zeta_2\geq 0$ and $\bm \Omega \succeq\bm 0$ are Lagrangian multipliers associated with \eqref{eq:proof_ob2_main_b}, \eqref{eq:proof_ob2_main_c} and $\bm \Phi \succeq \bm 0$, respectively. Then, part of the KKT conditions of problem~\eqref{eq:proof_ob2_main} are listed below:
\begin{subequations}\label{proof:ob2_kkt}
	\begin{align}
	\bm I + \zeta_2 \bm h(\theta_{2}) \bm h(\theta_{2})^H - \zeta_1  \bm h(\theta_{1}) \bm h(\theta_{1})^H & = \bm \Omega \label{proof:ob2_kkt_a} \\
	\bm \Omega \bm \Phi & = \bm 0 \label{proof:ob2_kkt_b}
	\end{align}
\end{subequations}     
Multiply both sides of \eqref{proof:ob2_kkt_a}  by $\bm \Phi$ and make use of \eqref{proof:ob2_kkt_b}  to get
\[ (\bm I + \zeta_2 \bm h(\theta_{2}) \bm h(\theta_{2})^H ) \bm \Phi = \zeta_1 \bm h(\theta_{1}) \bm h(\theta_{1})^H \bm \Phi \]
which implies that 
\begin{align*}
{\rm rank}(\bm \Phi) & =  {\rm rank}((\bm I + \zeta_2 \bm h(\theta_{2}) \bm h(\theta_{2})^H ) \bm \Phi) \\
&  =  {\rm rank} ( \zeta_1 \bm h(\theta_{1}) \bm h(\theta_{1})^H \bm \Phi ) \\
&  \leq 1,
\end{align*}
where the first equality is due to $\bm I + \zeta_2 \bm h(\theta_{2}) \bm h(\theta_{2})^H \succ \bm 0$. Since $\bm \Phi=\bm 0$ is infeasible for problem~\eqref{eq:proof_ob2_main}, we must have ${\rm rank}(\bm \Phi)=1$.

{\it Step 3:} Without loss of optimality, we can express the optimal $\bm \Phi^\star$ for problem~\eqref{eq:main_AN} as
\[ \bm \Phi^\star =  \bm y \bm y^H, ~~  \bm y = [\bm u_1, \bm u_2, \bm U^\perp] \bm x \]
for some $\bm x \in \mathbb{C}^N$ such that $\| \bm y \|=1$, where  $\bm U^{\perp} \in \mathbb{C}^{N\times (N-2)}$ is the orthogonal complement of $[\bm u_1 ~\bm u_2]$. Notice that $\bm y$ affects problem~\eqref{eq:main_AN} through $\bm h^H(\theta_{1}) \bm y$,  $\bm h^H(\theta_{2}) \bm y$, and that $[\bm u_1, \bm u_2]$ spans the same subspace as $[\bm h(\theta_{1}), \bm h(\theta_{2})]$. Any component of $\bm y$ in $\bm U^\perp$ has no contribution to problem~\eqref{eq:main_AN}. Therefore, without loss of optimality, $\bm y$ can be further expressed as
\[ \bm y =  x_1 \bm u_1 + x_2 \bm u_2.\]
Since $\bm u_2$  is orthogonal to $\bm u_1$ (or $\bm h(\theta_{1})$), $x_2\bm u_2$ affects only the legitimate communication user's SINR in an adversary manner. Clearly, at the optimal $\bm \varphi^\star$, we should set $x_2 =  0$ to save the power, which leads to $x_1 =1$ because of $\| \bm u_1 \|=1$.

\subsection{Proof of Theorem~\ref{lemma:drb}} \label{app:B}
The key to the proof of Theorem~\ref{lemma:drb} is to relate the distributionally robust constraint with Conditional Value-at-Risk (CVaR) functional, which is given by
\begin{equation} \label{eq:CVaR_def}
{\rm CVaR}_{\epsilon} (f(\bm x)) = \inf_{\nu \in \mathbb{R}} \left \{ \nu +  {\epsilon}^{-1} \mathbb{E}_{\cal P} [(f(\bm x) - \nu)^+] \right\}.
\end{equation} 
By noting that $f(\bm x)$ is quadratic in $\bm x$,  it follows from~\cite[Theorem 2.2]{Zymler} that the distributionally robust constraint~\eqref{eq:lemma_drb_original} is equivalent to
\begin{equation}\label{eq:proof_cvar}
\sup_{{\cal P} \in \mathscr{D}({\bar{\bm \mu}}, \bm \Sigma )} {\rm CVaR}_{\epsilon} (f(\bm x)) \leq 0.
\end{equation}
According to the definition of ${\rm CVaR}_{\epsilon}$, we have
\begin{subequations} \label{eq:eqv_CVaR_dual}
	\begin{align}
	&\sup_{{\cal P} \in \mathscr{D}({\bar{\bm \mu}}, \bm \Sigma )}  {\rm CVaR}_{\epsilon}(f(\bm x)) \label{eq:eqv_CVaR_dual_a} \\
	& = \sup_{{\cal P} \in \mathscr{D}({\bar{\bm \mu}}, \bm \Sigma )}\inf_{\nu \in \mathbb{R}} \Big \{\nu + \frac{1}{\epsilon} \mathbb{E}_{ {\cal P}} [(f(\bm x) - \nu)^+] \Big\}  \label{eq:eqv_CVaR_dual_b} \\
	& = \inf_{\nu\in \mathbb{R}} \Big \{ \nu + \frac{1}{\epsilon}  \sup_{{\cal P} \in \mathscr{D}({\bar{\bm \mu}}, \bm \Sigma )} \mathbb{E}_{\cal P} [(f(\bm x) - \nu)^+]  \Big \}. \label{eq:eqv_CVaR_dual_c}
	\end{align}
\end{subequations}
In~\eqref{eq:eqv_CVaR_dual_c}, we have interchanged the maximization and minimization operations, which can be justified by a stochastic saddle point theorem due to Shapiro and Kleywegt~\cite{Shapiro}. To express the supremum in~\eqref{eq:eqv_CVaR_dual_c} into a more tractable form, we need the following lemma:
\begin{Lemma} [ {\cite[Lemma A.1]{Zymler}} ] \label{lemma:2}
	Let $g:\mathbb{C}^N \rightarrow \mathbb{R}$ be a measurable function, and define the worst-case expectation $\theta_{\rm wc}$ as
	\[ \theta_{\rm wc} = \sup_{{\cal P} \in \mathscr{D}(  \bar{\bm \mu}, \bm \Sigma )} \mathbb{E}_{ \cal P} [(g(\bm x))^+].\]
	Then,
	\begin{align*}
	\theta_{\rm wc}  =\inf_{\bm Q \in \mathbb{H}^{N+1}}  &   ~ {\rm Tr}(\bm \Omega \bm Q) \\
	{\rm s.t.}&  ~~~ [\bm x^H ~1] \bm Q [\bm x^H ~ 1]^H \geq g(\bm x), ~\forall ~\bm x \in \mathbb{C}^N, \\
	& ~~~ \bm Q \succeq \bm 0
	%& \hspace{40pt} \bm M \succeq \bm 0,
	\end{align*}
	where
	$  \bm \Omega =
	\begin{bmatrix}
	\bm \Sigma +  \bar{\bm \mu}  \bar{\bm \mu}^H &  \bar{\bm \mu} \\
	\bar{\bm \mu}^H & 1
	\end{bmatrix}.$
\end{Lemma}

It follows from Lemma~\ref{lemma:2} that the supremum in \eqref{eq:eqv_CVaR_dual} is equal to
\begin{subequations} \label{eq:sup_value}
	\begin{align}
	\inf_{\bm Q \in \mathbb{H}^{N+1},}& ~~ ~ {\rm Tr}(\bm \Omega \bm Q) \label{eq:sup_value_a} \\
	{\rm s.t.} 	&   ~~~ [\bm x^H ~1] \bm Q [\bm x^H, ~ 1]^H \geq f(\bm x) - \nu, ~\forall \bm x , \label{eq:sup_value_b} \\
	& ~~~  \bm Q \succeq \bm 0.
	%& \hspace{15pt} ~ \bm M_i \succeq \bm 0, \label{eq:sup_value_c}
	\end{align}
\end{subequations}
Since $f(\bm x)$ is quadratic in $\bm x$, the quadratic constraint in~\eqref{eq:sup_value_b} holds for all $\bm x$ if and only if
\begin{equation}\label{eq:Q_expression}
\bm Q \succeq \begin{bmatrix}
\bm A & \bm b \\
\bm b^H & c - \nu
\end{bmatrix}.
\end{equation}
By replacing~\eqref{eq:sup_value_b} with~\eqref{eq:Q_expression} and substituting~\eqref{eq:sup_value} into~\eqref{eq:eqv_CVaR_dual_c}, we can re-express \eqref{eq:proof_cvar} equivalently as
\begin{subequations}\label{eq:conic_reformulation}
	\begin{align}
	& 0 \geq \inf_{\nu \in \mathbb{R},\bm Q\in \mathbb{H}^{N+1}} \nu + \epsilon^{-1} \text{Tr}(\bm \Omega \bm Q ) \label{eq:conic_reformulation_a} \\
	& \hspace{35pt}  {\rm s.t.}~~~ \eqref{eq:Q_expression}, \quad \bm Q \succeq \bm 0. \label{eq:conic_reformulation_b}
	\end{align}
\end{subequations}
It is easy to see that~\eqref{eq:conic_reformulation_a} holds if and only if there exists a  point $(\nu,\bm Q)$ satisfying~\eqref{eq:conic_reformulation_b} such that $\nu  + \epsilon^{-1} \text{Tr}(\bm \Omega \bm Q ) \leq 0$ holds. This completes the proof.

%\section{Data Availability}
%The data used to support the findings of this study are available from the corresponding author upon request.
%
%
%\section{Conflicts of Interest}
%The authors declare that they have no conflicts of interest.
%
%\section{Acknowledgments}
%This work was supported by the National Natural Science
%Foundation of China under Grant 62171110.

\end{document}